\documentclass[prd,aps,superscriptaddress,preprintnumbers,eqsecnum,nofootinbib,notitlepage]{revtex4-2}
\usepackage[utf8]{inputenc}
\usepackage{amsfonts,amsmath,amssymb,bm,subfigure}
\usepackage{multirow,hyperref}
\usepackage{xcolor}
\usepackage{comment}
\usepackage{graphicx}
\usepackage{orcidlink}

\begin{document}
\baselineskip=12pt

\title{$G_{3}$ - interacting scalar tensor dark energy}
\author{Masroor C. Pookkillath\orcidlink{0000-0002-7199-8037}}
\email{masroor.cha@mahidol.ac.th} 
\affiliation{Centre for Theoretical Physics and Natural Philosophy, Mahidol University, Nakhonsawan Campus,  Phayuha Khiri, Nakhonsawan 60130, Thailand}

\author{Nandan Roy\orcidlink{0000-0001-7197-453X}}
\email{nandan.roy@mahidol.ac.th} 
\affiliation{Centre for Theoretical Physics and Natural Philosophy, Mahidol University, Nakhonsawan Campus,  Phayuha Khiri, Nakhonsawan 60130, Thailand}

\begin{abstract}
We study the effect of adding an interaction in the $G_3$ term of Horndeski theory, where the propagation of gravitational waves are not modified. We derive the background and perturbation equations of motion from the action. We also derive the no-ghost and Laplacian instability conditions for tensor modes and scalar mode propagation. Then we study the evolution of the matter perturbation in the quasi-static approximation. We find that the gravitational couplings to the baryonic and cold dark matter over density are modified in this theory. We introduce a concrete model of the free function in the theory and study the background and linear perturbation dynamics. We then use the genetic algorithm to test the model. We compare the $H(z)$ function of the model and the $H(z)$ curve predicted by the genetic algorithm, using the $H(z)$ data. For the perturbation sector we compute the $f\sigma_{8}$ observable  for the model and compare it with the predicted function from the genetic algorithm from the $f\sigma_{8}$ data.
\end{abstract}

\maketitle

\section{Introduction}

In the $\Lambda$CDM description of the universe, the cosmological constant $\Lambda$ is introduced to the Einstein-Hilbert action to explain the observed cosmic expansion today and it has a constant equation of state $w_{\Lambda} = -1$. In addition, a fluid component with zero pressure (zero temperature, hence cold) is considered to explain the cold dark matter component. Even though the $\Lambda$CDM model is favored by existing cosmological data~\cite{SupernovaSearchTeam:1998fmf, SupernovaCosmologyProject:1998vns, Boomerang:2000efg, WMAP:2003elm, WMAP:2008rhx, Planck:2013pxb, Planck:2018vyg}, 
it fails to fully explain the cosmic expansion in late-time cosmology, the dark energy problem, and to provide an adequate description for cold dark matter. Additionally, the model struggles with several theoretical challenges, such as the cosmological constant problem, fine-tuning problem, and more. On top of the theoretical incompleteness of the $\Lambda$CDM model, there are growing tensions that began to appear after the 2013 release of data from the Planck mission~\cite{Planck:2013pxb}.

The value of the Hubble expansion rate today, $H_{0}$ inferred from the Planck Cosmic Microwave Background 2013 data is reported to be significantly lower than the value measured from the Type Ia supernovae~\cite{Riess:2011yx}. Over the years, this tension keeps increasing. Currently, the Plank$-2018$ result shows that the value of the current acceleration rate of the universe is $67.4\pm 0.5 \text{km} \text{s}^{-1} \text{Mpc}^{-1}$~\cite{Planck:2018vyg}. But from the local measurement from the SH0ES Collaboration, it is reported that the value of the present accelerated expansion is $74.1 \pm 1 \text{km} \text{s}^{-1} \text{Mpc}^{-1}$\cite{Riess:2021jrx}. This adds up to a tension of around $5 \sigma$ and is called the Hubble tension.

Another noticeable tension or discrepancy is in the estimation of the parameter $\sigma_{8} (S_{8})$, which measures the density of the fluctuation, that resulted in the formation of the structures that we see today. This parameter estimated from the cosmic microwave background radiation~\cite{Planck:2018vyg}, is higher than that measured from the Large scale structure data~\cite{DES:2021bvc}, which results in a $3\sigma$ discrepancy. These tensions in the estimation of major cosmological parameters from different kinds of observations suggest that the physics of the dark sectors of the universe might be different from the $\Lambda$CDM model.

The dynamic nature of dark energy is one of the alternatives to the cosmological constant which can help us to alleviate the current tensions. Modified theories of gravity are successful in giving a dynamical nature to the dark energy and dark matter phenomenon. For example, the quintessence model is introduced to explain the late-time cosmic expansion with a scalar field, instead of trivial cosmological constant $\Lambda$~\cite{Tsujikawa:2013fta}\footnote{There are also other models that can mimic the dark energy phenomenon, with out introducing any additional degrees of freedom~\cite{Lin:2017oow, Mukohyama:2019unx, DeFelice:2020eju, DeFelice:2020cpt,Aoki:2020oqc, Afshordi:2006ad, Mylova:2023ddj}.}. Later this model was generalised to more general theories of scalar-tensor gravity, like $k-$essence~\cite{Armendariz-Picon:2000ulo}, Horndeski gravity~\cite{Horndeski:1974wa}, beyond Horndeski~\cite{Gleyzes:2014dya} and Degenerate Higher Order Scalar Tensor (DHOST) theories~\cite{Langlois:2015cwa}. Vector-tensor theory is also introduced to explain the dynamical dark energy, for example~\cite{Armendariz-Picon:2004say, Tasinato:2014eka}. Vector-tensor dark energy was also generalised to a Generalised Proca theory~\cite{Heisenberg:2014rta}, further extended to Proca-Nuevo~\cite{deRham:2020yet}. The cosmology of vector-tensor theories are studied in~\cite{DeFelice:2016yws, DeFelice:2020sdq, deRham:2021efp}. The dark energy models with scalar field and the vector field were extended to interaction with cold dark matter. In the literature, the interaction is usually introduced with phenomenological approach. In~\cite{Pourtsidou:2013nha} a general Lagrangian for the interacting scalar field dark energy and dark matter is introduced, in which an interaction term is $\partial_{\mu} \phi u^{\mu}$, where $u^{\mu}$ is the four velocity of the cold dark matter. This interaction is then further extended to generalised scalar-tensor theory~\cite{Kase:2020hst}. For the dark energy modelled by vector field $A_{\mu}$, a similar interaction term, $Z = A_{\mu}u^{\mu}$ was introduced in~\cite{DeFelice:2020icf}. Using similar methods of~\cite{Pourtsidou:2013nha}, a generalised interacting Lagrangian with momentum transfer is studied~\cite{Pookkillath:2024ycd}, where new interaction terms were found.

Though modified theories of gravity are a promising approach to reducing the above-mentioned tensions~\cite{DiValentino:2021izs}. It has been noticed that on reducing the Hubble tension either from the late-time perspective or from pre-recombination modification the value of $S_{8}$ is worsened. This is one of the challenges the modified gravity models face. Interacting dark energy is proposed to address this issue. With interacting dark energy and dark matter, it is possible to suppress the effective gravitational coupling $G_{\rm eff}$ of the cold dark matter~\cite{DiValentino:2019ffd, DeFelice:2020icf, BeltranJimenez:2021wbq, Pookkillath:2024ycd}.

In the work, we propose an interacting dark sector, in which the interaction term between CDM and dark energy is $Z \equiv u^{\mu} \partial_{\mu}\phi$. We introduce this term into the Lagrangian by $L_{\rm int} = G_{3}(\phi, X, Z)\nabla_{\mu}\partial^{\mu} \phi$. There are a few motivations for introducing this new interaction model. First, while constructing an interacting scalar dark energy the general Lagrangian is given as $L( n^{\mu}, u^{\mu}, \partial_{\mu} \phi, \phi )$, where $n^{\mu}$ is the density current. However, the exploration of the cosmological phenomenology is restricted to assuming a quintessence scalar field together with interaction term $f(Z)$, or the Horndeski theory with the same interaction term. However, it is possible to have interaction term with higher order derivative terms, which can give non-trivial phenomenology. Second, it is already known that in the Horndeski theory, there exist higher derivative terms which modify the effective gravitational coupling at the level of linear perturbation. In the $G_{3}$ Horndeski theory for some value of the parameters, the Integrated Sachs-Wolfe (ISW)-Galaxy cross correlation is found to have negative values for some regions of parameters~\cite{Renk:2016olm, Giacomello:2018jfi}. It is noticed that, while lowering the value of the $G_{{\rm eff}}$ the correlation turns to positive, but higher values of the $G_{\rm eff}$ can have better fit with the observations. Considering this fact, it is interesting to introduce features like an interaction term in the $G_{3}$ function of the Hodrneski theory to see, if the negative cross-correlations can be removed. However, as a first step, we need to make sure that this introduction does not introduce any instabilities. 

We derive the background and perturbation equations of motion from the action. We also derived the no-ghost and Laplacian stability conditions for tensor modes and scalar modes propagation. We also find the evolution of the matter perturbation in the quasi-static approximation. We find that the effective gravitational coupling is modified in this theory.

Since this model is theoretically viable, we used the machine learning algorithm to test its observational viability using the $H(z)$ data, and the redshift space distortion data. In particular, we use the genetic algorithm to study preliminary viability of this model, with the $H(z)$ data and $f \sigma_{8}$ data.

Genetic algorithm has been widely used in the context of astrophysics~\cite{Luo:2019qbk, Ho:2019zap} and cosmology~\cite{Bogdanos:2009ib, Nesseris:2010ep, Nesseris:2012tt, Nesseris:2013bia, Sapone:2014nna, Arjona:2019fwb}. It is a machine learning technique used for non-parametric reconstruction of data which  is motivated by evolutionary biology. This algorithm starts with a randomly generated initial population based on the predefined grammar of basis functions such as $\sin$, exp, log, etc., and the mathematical algebraic operations ``$+$, $-$, $\times$, $\div$ ''. Then the algorithm allows to generate the offspring and the fitness function is evaluated. In this context, the fitness function is the measure of how well the population describes the data, which is the measure of $\chi^2$. On each generation of successive populations, genetic operators like mutation and crossover are applied. This process ends once the desired criteria are reached.

Here is the basic working scheme of genetic algorithm
\begin{enumerate}
    \item Generate an initial random population of functions $M(0)$ based on a predefined grammar.
    \item Calculate the fitness for each individual in the current population $M(t)$.
    \item Create the next generation $M(t+1)$ by probabilistically choosing individuals from $M(t)$ to produce offspring through crossover and mutation, and possibly also keeping a proportion of the previous generation $M (t)$.
    \item Repeat step 2 until a termination goal has been achieved.
\end{enumerate}

When the next population is generated, the selection rate for the generation of the next offspring is typically $10\% - 20\%$. The mutation rate for the generation of the next offspring is $5\% \sim 10\%$. Mutation rate is the probability that an arbitrary part of an individual is changed. These two rates will affect the rate of convergence, and typically if we increase the rate more than what is specified, typically the algorithm will not converge. The major difference between the genetic algorithm and the standard analysis is that it does not need any model with a given number of parameters a priori. It only needs observational data; hence it can be used to test any given theory.

We can use the genetic algorithm to test a given model and see if this model can fit the observable within the error bar predicted by the genetic algorithm. We exploit this possibility to test the model we have introduced in this work. In particular, we use the $H(z)$ data and $f \sigma_{8}$ data compilation used in the~\cite{Alestas:2022gcg}. We also use the genetic algorithm code introduced in~\cite{Alestas:2022gcg}. 

This manuscript is arranged as follows. In the section~\ref{sec:thoery}, we introduce the interacting model with $G_{3}(\phi, X, Z)$, where $X\equiv -1/2\partial_{\mu}\phi \partial^{\mu}\phi$, and $Z \equiv \partial_{\mu}\phi u^{\mu}$. Then in the section~\ref{sec:background_eom}, we derive the equations of motion for the background cosmology assuming flat-Friedmann-Lema\^{i}tre-Robertson-Walker metric (FLRW) metric. In the section~\ref{sec:ghost_condition}, we derive the no-ghost condition and the Laplace instability conditions for the both tensor and the scalar modes. For the tensor modes, it is trivial. In the section~\ref{sec:perturbation_eom}, we derive the equations of motion for the linear perturbations. Subsequently, we study the evolution of the baryonic and cold dark matter over-densities in~\ref{sec:QSA_limit}. Then, we introduce a concrete model in section~\ref{sec:concrete_model} and study the background dynamics and matter over-density dynamics with a concrete model. In the section~\ref{sec:GA}, we compute the $H(z)$, and $f\sigma_{8}$ and use genetic algorithm and test the model using $H(z)$ and $f\sigma_{8}$ data compilation~\cite{Alestas:2022gcg} respectively. Finally, in section~\ref{sec:conclusion} we conclude our work.

\section{The theory}\label{sec:thoery}
In this section, we propose an interacting scalar field dark energy model. Usually, the interacting scalar field dark energy models are typically discussed either phenomenologically or by introducing a momentum transfer term, such as $\sqrt{-g} f \big(u_{c}^{\mu}\partial_{\mu}\phi \big)$, or by including conformal coupling in the Lagrangian~\cite{Clemson:2011an, Pourtsidou:2013nha, Kase:2020hst}. In this work, we use the second proposal by introducing a momentum transfer interaction in the kinetic braiding term by adding a term $\sqrt{-g} F\big( u_{c}^{\mu} \nabla_{\mu}\phi \big) \nabla_{\mu}\nabla^{\mu}\phi$ into the kinetic braiding scalar-tensor theory. We name this model as $G_{3}$ interacting scalar tensor theory since the kinetic braiding term is the $G_{3}$ - term of the Horndeski gravity~\cite{Horndeski:1974wa, DeFelice:2011bh}. Without loss of any generality one can consider the $G_{3}$ interaction term as two additive functions of the $X$ and $Z$ such as  $G_{3}(\phi, X, Z) \nabla_{\mu}\nabla^{\mu} \phi = G(X) \nabla_{\mu}\nabla^{\mu} \phi + F(Z) \nabla_{\mu}\nabla^{\mu} \phi$ . The action of the interacting scalar tensor dark energy can be written as
\begin{equation}
\label{eq:action}
S=\int{\rm d}^{4}x\sqrt{-g}\left[\frac{M_{{\rm P}}^{2}}{2}R - \alpha_{1} \big(X - V(\phi) \big) - G\left(X \right)\nabla_{\mu}\nabla^{\mu}\phi + F\left(Z \right)\nabla_{\mu}\nabla^{\mu}\phi\right] + S_{m} + S_{c}\,,
\end{equation}
where $g$ is the determinant of the metric tensor $g_{\mu\nu}$, and $X\equiv-\frac{1}{2}\partial_{\mu}\phi \partial^{\mu}\phi$. $Z$ is the interaction
term defined as 
\begin{equation}
Z\equiv u_{{\rm c}}^{\mu}\nabla_{\mu}\phi\,,
\end{equation}
where $u_{{\rm c}}^{\mu}$ is the four-velocity of the cold dark
matter component in the universe. The coefficient $\alpha_{1}$ is a dimensionless coefficient. The matter action for radiation and baryon is represented by $S_m$ and given by Schutz-Sorkin action~\cite{Schutz:1977df, Brown:1992kc, DeFelice:2009bx, Pookkillath:2019nkn},
\begin{equation}\label{eq:Sm}
S_{m}=-\sum_{{\rm m= b,r}}\int{\rm d}^{4}x\sqrt{-g}\left[\rho_{\rm m}\left(n_{\rm m}\right)+J_{\rm m}^{\mu}\left(\partial_{\mu}l_{\rm m}\right)\right]\,,
\end{equation}
where $n_{\rm m}$ is the number density and $J^{\mu}_{\rm m}$ is the current density, respectively defined as
\begin{equation}\label{eq:nDef}
n_{\rm m}=\sqrt{-J_{\rm m}^{\mu} \, J_{\rm m}^{\nu} \, g_{\mu\nu}}\,,
\end{equation}
and 
\begin{equation}\label{eq:Jdef}
    J_{\rm m}^{\mu} = n_{\rm m} u_{\rm m}^{\mu}\,.
\end{equation}
The dark matter action is given as
\begin{equation}
    S_{c} = -\int {\rm d}^4 x \sqrt{-g} \left[ \rho_{c} (\tilde{n}) + n_{c}^{\mu } \nabla_{\mu }K \right]\,,
\end{equation}
where, $\rho_{c}$ is the energy density and $K$ is Lagrangian multiplier to impose the number density conservation.
\begin{equation}
    \tilde{n} = -n_{{c}\mu}n_{c}{}^{\mu}\,.
\end{equation}
This action is explored in~\cite{Pookkillath:2024ycd} in the context of interacting vector dark energy. 

\section{Background equations of motion} \label{sec:background_eom}
We consider homogeneous and isotropic flat-Friedmann-Lema\^{i}tre-Robertson-Walker (FLRW) metric with scalar perturbations 
\begin{equation}
{\rm ds}^{2}=-\left(1+2\alpha\right){\rm d}t^{2}+2 \partial_{i}\chi{\rm d}\tau{\rm d}x^{i}+a^{2}\left[\left(1+2\zeta\right)\delta_{ij}+\partial_{i}\partial_{j}E/a^{2}\right]{\rm d}x^{i}{\rm d}x\,.
\end{equation}
For the fields on the matter action we define
\begin{equation}\label{eq:JmuDef}
      J_{\rm m}^{\nu} = \left[\frac{J_{\rm m}}{a}\left(1+\delta J_{\rm m}\right),\frac{1}{a^{2}}\delta^{ij}\partial_{j}\delta j_{\rm m}\right]\,, 
\end{equation}
and
\begin{equation}
    l_{\rm m} = \bar{l}_{\rm m}+\delta l_{\rm m}\,.   
\end{equation}
We define the four-velocity of the cold dark matter component
\begin{equation}\label{eq:umuDef}
    u^{\mu}_{\rm c} = \frac{n_{\rm c}^{\mu}}{\sqrt{\tilde{n}}} \,,
\end{equation}
where 
\begin{equation}
  n_{\rm c }^{\mu} \equiv \left[ \Bar{n}_{0}(1+ \delta n_{0}), \frac{1}{a^2} \delta^{il} \partial_{l} n_{s } \right]  \,,
\end{equation}
then the four-velocity of the cold dark matter component automatically satisfies
\begin{equation}
    u_{\rm c}^{\mu}u_{\mu}{}_{\rm c} = -1 \,,
\end{equation}
and we can define the Lagrangian multiplier 
\begin{equation}
    K = K_{0} + \delta K\,.
\end{equation}

The variation with respect to the field $g_{\mu\nu}$ gives the following Friedmann equations of motion
\begin{equation}
3M_{{\rm p}}^{2}H^{2}=\sum_{I={\rm c,b,r}}\rho_{I}+\rho_{{\rm DE}}\,, \label{eq:1stFE}
\end{equation}
where the density of the dark energy is given by
\begin{equation}
    \rho_{{\rm DE}} \equiv   - \alpha \frac{1}{2} \dot{\phi}^2 - \alpha V(\phi) + 3H \dot{\phi}^2 F_{,Z} - 3 H G_{,X} \dot{\phi}^3 \,.
\end{equation}
and the second Friedmann equation can be written as
\begin{equation}
2 M_{\rm P}^2 \frac{\dot{H}}{a} = \sum_{I={\rm c,b,r}}  \left(\rho_{I} + P_{I} \right) + \rho_{{\rm DE}} + P_{{\rm DE}}\,, \label{eq:2ndFE}
\end{equation}
where the pressure of the dark energy is given by
\begin{equation}
    P_{\rm DE} = - \alpha\frac{1}{2}\dot{\phi}^2 + \alpha \, V(\phi) + F_{, Z} \dot{\phi} \ddot{\phi} - G_{,Z} \dot{\phi}^2 \ddot{\phi} \,.
\end{equation}
The equation of state for dark energy is defined as 
\begin{equation}
w_{{\rm DE}}\equiv\frac{P_{{\rm DE}}}{\rho_{{\rm DE}}}\,.
\end{equation}

Now we can also derive the equation of motion for the field $\phi$
\begin{eqnarray}
    & & \ddot{\phi} \left[ \alpha + 3 H F_{,Z} + \left(3 H\dot{\phi} F_{,ZZ} - 6 H G_{,X}  \right) \dot{\phi} - 3 H \dot{\phi}^3  G_{,XX} \right] + \left( -9 H^2 G_{,X} - 3 \dot{H} G_{,X} \right) \dot{\phi}^2 \nonumber \\
    & & + \left( \alpha 3 H + 9 H^2 F_{,Z} + 3 F_{,Z} \dot{H}  \right) \dot{\phi} - V(\phi) = 0 \,. \label{eq:eqdphiBG}
\end{eqnarray}

In this section, we have derived the background equations of motion for the FLRW metric. We proceed to study the no-ghost conditions and perturbation equations of motion in the following sections.

\section{ghost Conditions} \label{sec:ghost_condition}
In this section, we show the no-ghost and Laplacian instability conditions for both the tensor modes and the scalar propagating modes for the model given by the action in Eq.~(\ref{eq:action}). Since we are proposing a new interacting dark energy model, it is unavoidable to perform those analysis.
\subsection{Tensor modes}
Let us first explore the propagation of the tensor modes. Considering the tensor mode perturbation around the flat-FLRW background,
\begin{equation}
\label{eq:metric_tensor}
    {\rm d} s^2 = \left[-{\rm d}t^2 + a(t)^2  \left(\delta_{ij} + \sum_{\lambda = +, \times} \epsilon_{ij}^{\lambda} h_{\lambda} \right) {\rm d}x^{i}{\rm d}x^{j} \right] \,,
\end{equation}
where we choose the metric in such a way that the tensor mode perturbations obey standard trace-less condition $\delta_{ij}\epsilon_{ij} = 0 $, and divergence free gauge condition $\epsilon_{ij}^{\lambda} \delta^{jl} \partial_{l}h_{\lambda} = 0$. It also obeys the normalisation condition $\epsilon_{ij}^{+} \epsilon_{kl}^{+} \delta^{ik} \delta^{jl} = \epsilon_{ij}^{\times} \epsilon_{kl}^{\times} \delta^{ik} \delta^{jl} = 1$, with $\epsilon_{ij}^{+}\epsilon_{kl}^{\times}\delta^{ik} \delta^{jl} = 0  $. Doing the standard procedure of expanding the action Eq.(\ref{eq:action}) with the metric Eq.(\ref{eq:metric_tensor}) up to second order, and substituting the background equations of motion we get the following reduced action for the tensor modes
\begin{equation}
    S_{\rm T} = \frac{M_{\rm P}^2}{4} \sum_{\lambda = +, \times} \int {\rm d}^4 x \, a^3 \left[ \dot{h}_{\lambda}^{2} - \frac{k^2}{a^2} h_{\lambda}^{2} \right] \,,
\end{equation}
notice that the action has been changed into Fourier space, and $k$ is the Fourier mode. Therefore, it shows that the tensor mode propagation does not change in this theory. 

\subsection{Scalar Modes}
Now let us discuss the stability conditions for scalar mode propagation. To make the calculations easy, we choose a unitary gauge, that is, we keep the fields $\delta \phi = 0$, and $E = 0$, to fix the gauge freedom. We also make the field redefinition as follows
\begin{eqnarray}
    \delta J_{\rm m} = \frac{N_{\rm m}}{a^3}\left( \frac{\rho}{n\rho_{,n}} \delta_{\rm m } - \alpha \right) \,, \\
    \delta n_{0} = \frac{N_{\rm c}}{a^3} \left( \frac{\rho_{c}}{2 \tilde{n} \rho_{c,\tilde{n}}} \delta_{\rm c} - \alpha \right) \,,
\end{eqnarray}
then we integrate out the variables $\delta j_{\rm m}$, $n_{s}$, $\delta_{\rm m} $, $\delta_{\rm c}$. Now we are remaining with the variables $\delta l_{\rm m}$, $\delta l_{\rm c}$, and $\chi$, $\zeta$ and $\chi$. Varying the field $\chi$, we get the field equation
\begin{equation}
    \left[2 M_{\rm P}^{2} H + F_{,Z} \dot{\phi}^2 - G_{,X} \dot{\phi}^3 \right]\alpha = 2 M_{\rm P}^{2} \dot{\zeta} - n_{\rm m} \delta l_{\rm m} - n_{\rm c} \delta l_{\rm c} \,.
\end{equation}
We can use the above equation to replace the $\alpha$ field. Then, the fields $\chi$ and $\alpha$ are removed from the action. The remaining fields are $\zeta$,  $\delta l_{\rm m}$, and $\delta l_{\rm c}$, which are dynamical. 

After making a few integration by part one can write the action in the form
\begin{equation}
    \int {\rm d}^3x {\rm d}t \, a^3 \left[ \dot{\mathcal{U}}^{T}\, \mathbf{K}\, \dot{\mathcal{U}} -\frac{k^2}{a^2} \mathcal{U}^{T}\, \mathbf{G} \, \mathcal{U} - \mathcal{U}^{T}\, \mathbf{M} \, \mathcal{U} + \mathcal{U}^{T}\, \mathbf{B} \, \dot{\mathcal{U}} \right] \,, \label{eq:can_action}
\end{equation}
where $\mathbf{K}$, $\mathbf{G}$, $\mathbf{M}$, and $\mathbf{B}$ are 3$\times$3 matrix and the vector $\mathcal{U}^{T}$ is given by 
\begin{equation}
    \mathcal{U}^{T} \equiv (\zeta, \delta l_{\rm c}, \delta l_{m})\,.
\end{equation}

From the kinetic matrix, we have the following ghost conditions for the independent modes that are propagating, namely, the cold dark matter component, the matter component, consisting of relativistic particles (radiation), and non-relativistic particles (baryons), and the scalar mode
\begin{eqnarray}
    Q_{\rm c} & = & (\rho_{\rm c} + P_{\rm c}) >0 \, , \\
    Q_{\rm m} & = & (\rho_{\rm m} + P_{\rm m}) >0 \, \qquad {\rm m = r, b} \,, \\
    Q_{s} & = & \frac{M_{\rm P}^2 \dot{\phi} ^2 \left[-2 \alpha_{1} M_{\rm P}^2-6 \gamma  \dot{\phi} ^3 F_{,Z} G_{,X}+3 \gamma ^2 \dot{\phi} ^2 F_{,Z}^2-6 M_{\rm P}^2 H  \left\{\dot{\phi}  \left(\gamma  F_{,ZZ} - \dot{\phi} ^2 G_{,XX} - 2 G_{,X}\right)+\gamma  F_{,Z}\right\}+3 \dot{\phi} ^4 G_{,X}^2\right]}{\left[\dot{\phi} ^2 \left(\gamma  F_{,Z}-\dot{\phi}  G_{,X}\right)+2 M_{\rm P}^2 H \right]^2} >0\,. \nonumber \\
\end{eqnarray}
In the limit of no interaction we can recover the results of~\cite{DeFelice:2011bh}.

To find the speed of the propagation for each modes we can solve the following equation
\begin{equation}
    \text{det}(c_{s}^{2} \mathbf{K} - \mathbf{G}) = 0 \,.
\end{equation}
Since the above equation is cubic in $c_{s}^{2}$, we will have three solutions, which corresponds to the speed of the propagation of the three (four considering the matter to be both baryon and radiation) independent propagating modes. 

The speed of propagation of the matter sector composed of radiation and baryons we get
\begin{equation}
    c_{s}^{2}{}_{\rm m}  = \frac{\delta P_{\rm m}}{\delta \rho_{\rm m}} \, .
\end{equation}
For the cold dark matter sector since, we assume pressure and its perturbation is vanishing, we find that 
\begin{equation}
    c_{s}^{2}{}_{\rm c}  = 0 \,.
\end{equation}
Finally, we have one more solution which corresponds to the propagation of the scalar mode 
\begin{equation}
    c_{s}^{2}{}_{\zeta} = \frac{\mathcal{A}}{\mathcal{B}} \,,
\end{equation}
where
\begin{eqnarray}
    \mathcal{A} & \equiv & \gamma  F_{,Z} \left(3 H  \dot{\phi} +\ddot{\phi} \right) \left[2 M_{\rm P}^2 \dot{\phi}  \left(\alpha_{1}+\gamma  \ddot{\phi}  F_{,ZZ}+4 \gamma  H  F_{,Z}-2 \ddot{\phi}  G_{,X}\right)+\dot{\phi} ^3 \left(\gamma ^2 F_{,Z}^2-2 M_{\rm P}^2 \ddot{\phi}  G_{,XX}\right) \right. \nonumber \\
    & & \left. -2 \gamma  \dot{\phi} ^4 F_{,Z} G_{,X}+2 \gamma  M_{\rm P}^2 \ddot{\phi}  F_{,Z}-8 M_{\rm P}^2 H  \dot{\phi} ^2 G_{,X}+\dot{\phi} ^5 G_{,X}^2\right]  -\rho_{\rm c}  \left[2 \alpha_{1} M_{\rm P}^2+2 \gamma  M_{\rm P}^2 \ddot{\phi}  F_{,ZZ}\right. \nonumber \\
    & & \left.+2 M_{\rm P}^2 H  \left(\gamma  F_{,Z}-4 \dot{\phi}  G_{,X}\right) - 2 \gamma  \dot{\phi} ^3 F_{,Z} G_{,X}+\gamma ^2 \dot{\phi} ^2 F_{,Z}^2-2 M_{\rm P}^2 \dot{\phi} ^2 \ddot{\phi}  G_{,XX}-4 M_{\rm P}^2 \ddot{\phi}  G_{,X}+\dot{\phi} ^4 G_{,X}^2\right] \,,
\end{eqnarray}
and 
\begin{eqnarray}
    \mathcal{B} & \equiv & \left[\gamma  \dot{\phi}  F_{,Z} \left(3 H  \dot{\phi} +\ddot{\phi} \right)-\rho_{\rm c} \right] \times \nonumber \\
    & & \left[2 \alpha_{1} M_{\rm P}^2+6 \gamma  \dot{\phi} ^3 F_{,Z} G_{,X}-3 \gamma ^2 \dot{\phi} ^2 F_{,Z}^2+6 M_{\rm P}^2 H  \left\{\dot{\phi} \left(\gamma  F_{,ZZ} - \dot{\phi} ^2G_{,XX} - 2 G_{,X}\right)+\gamma F_{,Z}\right\}-3 \dot{\phi} ^4 G_{,X}^2\right] \,. \nonumber \\
\end{eqnarray}
In the limit of no interaction, we recover the results of~\cite{DeFelice:2011bh}. For the avoidance of the Laplacian instability we need the condition that $c_{s}^{2}{}_{\zeta} = \mathcal{A}/ \mathcal{B} >0$. 

\section{Perturbation equations} \label{sec:perturbation_eom}
Until now, we have explored the cosmological background, no-ghost and Laplacian instability conditions for the propagating scalar and tensor modes. In this section, we derive the equations of motion for the linear perturbation. Then, we study evolution of both baryonic and cold dark matter component in the quasi-static approximation limit.

We expand the action up to second order and varying the field $j_{\rm m}$ and $n_{s}$, the fields defined in Eq.~(\ref{eq:JmuDef}) and  Eq.~(\ref{eq:umuDef}) we get the following constraint equations respectively
\begin{eqnarray}
    \delta l & = &  \rho_{,n_{\rm m}} \chi + \frac{\rho_{,n_{\rm m}}}{n_{\rm m}} j_{\rm m} \, , \\
    \delta K & = & \frac{a^3}{N_{c}}\left( - 3 H F_{,Z} \dot{\phi} - F_{,Z} \ddot{\phi} \right)  \delta \phi + \left[ \frac{2 N_{c}}{a^3} \rho_{c , \tilde{n}} + \frac{a^3}{N_{c}} \dot{\phi}\left( - 3 H F_{,Z} \dot{\phi} - F_{,Z} \ddot{\phi} \right)\right] \chi \nonumber \\
    & & + \left[ 2 \rho_{c, \tilde{n}} + \frac{a^6}{N_{c}^2} \dot{\phi}\left( - 3 H F_{,Z} \dot{\phi} - F_{,Z} \ddot{\phi}\right) \right] n_{s}\,,
\end{eqnarray}
these equations can be used to  replace $\delta l$ and $\delta K$ and we also make another field redefinition
\begin{eqnarray}
    j_{\rm m} & = & -\frac{N_{\rm m}}{a^3} \frac{a}{k^2}\theta_{\rm m} \,, \\
    n_{s} & = & - \frac{N_{\rm cmd}}{a^3} \frac{a}{k^2}\theta_{c}\,.
\end{eqnarray}

The second order action in Newtonian Gauge
\begin{equation}
    \alpha = \Psi, \qquad \zeta = - \Phi, \qquad \chi = 0, \qquad \text{and} \qquad E = 0 \,,
\end{equation}
can be written as 
\begin{equation}
    S^{(2)} = \int \sqrt{-g} d^{4}x a^{3} \left[L^{(2)}_{R} + L^{(2)}_{\phi} + L^{(2)}_{Z}  + L^{(2)}_{\rm m} + L^{(2)}_{c} \right] \,,
\end{equation}
\begin{eqnarray}
    L^{(2)}_{R} & \equiv & M_{\rm P}^{2} \left[ \frac{k^2}{a^2} \Phi  (\Phi -2 \Psi )-3 \left( H ^2 \Psi ^2 -\Phi ^2 \dot{H} +2 H  \Psi  \dot{\Phi} +\dot{\Phi} ^2\right) \right] \,, \\
    & & \nonumber \\
     L^{(2)}_{\phi} & \equiv & \alpha_{1} \left[ \frac{k^2}{a^2}\frac{ \delta \phi ^2}{2 } + \delta \dot{\phi}  \left(3 \Phi  \dot{\phi} + \Psi  \dot{\phi} \right)- \frac{1}{2} \delta \dot{\phi}^2 -  \frac{3}{2} \Phi ^2 \dot{\phi} ^2 - \frac{1}{2} \Psi ^2 \dot{\phi} ^2 \right] - \frac{k^2}{2a^2}  \delta \phi ^2 \left[4 H  \dot{\phi} G_{X}+\ddot{\phi}  \left(\dot{\phi} ^2 G_{,XX}+2 G_{X}\right)\right]  \nonumber \\
    & & +\delta \dot{\phi}  \left[-9 H  \Phi  \dot{\phi} ^2 G_{X}-3 \dot{\phi} ^2 \left\{\dot{\Phi}  G_{X}+H  \Psi  \left(\dot{\phi} ^2 G_{,XX}+3 G_{X}\right)\right\}\right]+\frac{3}{2} H  \delta \dot{\phi} ^2 \dot{\phi}  \left(\dot{\phi} ^2 G_{,XX}+2 G_{X}\right)  \nonumber \\
    & & +\frac{3}{2} \Phi ^2 \dot{\phi} ^2 G_{X} \left(3 H  \dot{\phi} -\ddot{\phi} \right)+\frac{3}{2} \Psi  \dot{\phi} ^3 \left[2 \dot{\Phi} G_{X}+H  \Psi  \left(\dot{\phi} ^2 G_{,XX}+4 G_{X}\right)\right] -\frac{k^2}{a^2} \delta \phi  \Psi  \dot{\phi} ^2 G_{X} \,, \\
    & & \nonumber \\
    L^{(2)}_{F} & \equiv & \theta_{c}  \left[\frac{a}{k^2}  \dot{\delta}_{c}  \dot{\phi}  F_{,Z} \left(3 H  \dot{\phi} +\ddot{\phi}\right) - \frac{a}{k^2} 3 \dot{\Phi}  \dot{\phi}  F_{,Z} \left(3 H  \dot{\phi} +\ddot{\phi}\right) \right]  -\frac{3}{2} \Psi  \dot{\phi} ^2 \left[2 \dot{\Phi}  F_{,Z} +H  \Psi  \left(\dot{\phi}  F_{,ZZ}+3 F_{,Z}\right)\right]  \nonumber\\
    & &+\delta \phi  \left[\frac{k^2}{a^2} \Psi  \dot{\phi}  F_{,Z} +\delta_{c}  \left\{\ddot{\phi}^2 F_{,ZZ}+F_{,Z}  \left(3 \dot{H}  \dot{\phi} +\tfrac{d^{3} \phi}{dt^{3}} \right)+9 H ^2 \dot{\phi}  F_{,Z}+3 H  \ddot{\phi} \left(\dot{\phi}  F_{,ZZ}+2 F_{,Z}\right)\right\} \right. \nonumber \\
    & & \left. -3 \Phi  \left\{\ddot{\phi}^2 F_{,ZZ}+F_{,Z} \left(3 \dot{H} \dot{\phi} +\tfrac{d^{3} \phi}{dt^{3}} \right)+9 H ^2 \dot{\phi}  F_{,Z}+3 H  \ddot{\phi} \left(\dot{\phi}  F_{,ZZ}+2 F_{,Z}\right)\right\}\right] +\frac{3}{2} \Phi ^2 \dot{\phi}  F_{,Z} \left(\ddot{\phi}-3 H  \dot{\phi} \right) \nonumber \\
    & & +\frac{k^2}{2a^2} \delta \phi ^2 \left(\ddot{\phi} F_{,ZZ}+H  F_{,Z}\right) +\delta \dot{\phi}  \left[\delta_{c}  F_{,Z} \left(3 H  \dot{\phi} +\ddot{\phi}\right)-3 \Phi  \ddot{\phi}F_{,Z}+3 \dot{\phi}  \left\{\dot{\Phi}  F_{,Z}+H  \Psi  \left(\dot{\phi}  F_{,ZZ}+2 F_{,Z}\right)\right\}\right] \nonumber \\
    & & -\frac{3}{2} H  \delta \dot{\phi} ^2 \left(\dot{\phi}  F_{,ZZ}+F_{,Z}\right)+\frac{1}{2k^2}\theta_{c} ^2 \dot{\phi}  F_{,Z} \left(3 H  \dot{\phi} +\ddot{\phi}\right) \,, \\
    & & \nonumber \\
    L^{(2)}_{\rm m} & \equiv & -\frac{a  \rho_{\rm m}  \theta_{\rm m}  \dot{\delta}_{\rm m} }{k^2}+\delta_{\rm m}  \left(\frac{3 a  H  \theta_{\rm m}  (\delta \rho_{\rm m}    P_{\rm m} -\delta P_{\rm m}  \rho_{\rm m} )}{k^2 \delta \rho_{\rm m} }-\Psi  \rho_{\rm m} \right)-\frac{\theta_{\rm m} (P_{\rm m} +\rho_{\rm m} ) \left(\theta_{\rm m} -6 a  \dot{\Phi} \right)}{2 k^2} \nonumber \\
    & & -\frac{\delta_{\rm m} ^2 \delta P_{\rm m}  \rho_{\rm m} ^2}{2 \delta \rho_{\rm m}  (P_{\rm m} +\rho_{\rm m} )}     +\frac{3}{2} \Phi ^2 (P_{\rm m} +\rho_{\rm m} ) \,,  \\
    & & \nonumber \\
    L^{(2)}_{c} & \equiv & \theta_{c}  \left(\frac{3 a \rho_{c}  \dot{\Phi} }{k^2}-\frac{a \rho_{c}  \dot{\delta}_{c} }{k^2}\right)-\delta_{c} \Psi \rho_{c} -\frac{\rho_{c}  \theta_{c} ^2}{2 k^2} +\frac{3}{2} \Phi ^2\rho_{c} \,.
\end{eqnarray}

\subsection{Matter sector}
On varying the fields $\theta_{\rm m}$ and $\delta_{\rm m}$ we get the following equations of motion
\begin{eqnarray}
        E_{\theta_{\rm m}} & \equiv &  \dot{\delta}_{\rm m} + 3 H \left( \frac{\delta P_{\rm m}}{\delta \rho_{\rm m}} - \frac{P_{\rm m}}{\rho_{\rm m}} \right) \delta_{\rm m} + \left(1 + \frac{P_{\rm m}}{\rho_{\rm m}} \right)\frac{\theta_{\rm m}}{a} - 3 \left( 1 + \frac{P_{\rm m}}{\rho_{\rm m}}  \right) \dot{\Phi} \,, \\
        E_{\delta_{\rm m}} & \equiv & \dot{\theta}_{\rm m} + H \left( 1 - 3 \frac{\delta P_{\rm m}}{\delta \rho_{\rm m}} \right) \theta_{\rm m} - \frac{k^2}{a} \left( \frac{\delta P_{\rm m}/\delta \rho_{\rm m}}{1+ P_{\rm m}/\rho_{\rm m}} \right)\delta_{\rm m} - \frac{k^2}{a} \Psi + k^{2}\sigma_{\rm m}\,.
\end{eqnarray}
These equations are the equations of motion of the standard matter fluids in the system we are considering, in particular, radiation and baryon. For the baryon, we have $P_{\rm b}=0$ and $\delta P_{\rm b}=0$. We can add an effective shear term to the matter action $S_{\rm shear}^{(2)} = \int dt d^{3}x a^{4} \sigma_{\rm m} \left[\delta \rho_{\rm m} + (\rho_{\rm m} + P_{\rm m}) \zeta \right]$~\cite{DeFelice:2020cpt}, such that we get the equations of motion with the shear term.
\subsection{Dark matter sector}
Now we find the following equations of motion for the dark matter fluid. Here, we can see the implications of the interaction we have considered. The evolution of cold dark matter over-density is
\begin{eqnarray}
    E_{\theta_{c}} & \equiv & \dot{\delta}_{c} + \frac{\theta_{c}}{a} - 3 \dot{\Phi} = 0 \,,
\end{eqnarray}
and the evolution of the velocity potential is
\begin{eqnarray}\label{EQ:EQdeltadm}
    E_{\delta_{c}} & \equiv & \frac{ k^2 \delta \dot{\phi}  F_{,Z} \left(3 H  \dot{\phi} +\ddot{\phi} \right)}{a  \left[\rho_{c} -   \dot{\phi} F_{,Z} \left(3 H  \dot{\phi} +\ddot{\phi} \right)\right]}+\frac{ k^2 \delta \phi  \left[\ddot{\phi} ^2 F_{,ZZ}+F_{,Z} \left(3 \dot{H}  \dot{\phi} +\tfrac{d^{3} \phi}{dt^{3}} \right)+9 H ^2 \dot{\phi}  F_{,Z}+3 H  \ddot{\phi}   \left(\dot{\phi}  F_{,ZZ}+2 F_{,Z}\right)\right]}{a \left[\rho_{c} -   \dot{\phi}  F_{,Z} \left(3 H  \dot{\phi} +\ddot{\phi} \right)\right]} \nonumber \\
    & & -\frac{\theta_{c}  \left[ \dot{\phi}  \ddot{\phi} ^2 F_{,ZZ}+  F_{,Z} \left(3 \dot{H}  \dot{\phi} ^2+\ddot{\phi} ^2+\tfrac{d^{3} \phi}{dt^{3}}  \dot{\phi} \right)+12    H ^2 \dot{\phi} ^2 F_{,Z}+   H  \dot{\phi}  \ddot{\phi}  \left(3 \dot{\phi}  F_{,ZZ}+10 F_{,Z}\right)-H \rho_{c} \right]}{\rho_{c} - \dot{\phi}  F_{,Z} \left(3 H  \dot{\phi} +\ddot{\phi} \right)} \nonumber \\
    & & -\frac{k^2 \psi  \rho_{c} }{a  \left[\rho_{c} -   \dot{\phi}  F_{,Z} \left(3 H  \dot{\phi} +\ddot{\phi} \right)\right]} +\dot{\theta}_{c} = 0 \,.
\end{eqnarray}

\subsection{Gravity Sector}

Here are the equations of motion for the gravity sector, including the scalar field equation of motion. The Hamiltonian constraint equation is 
\begin{eqnarray}\label{EQ:EQalpha}
    E_{\alpha} & \equiv & - \sum_{I = c, r, b} \delta_{I} \rho_{I} +  \left(3 H \dot{\phi}^2 F_{,ZZ} + 6 H \dot{\phi}  F_{,Z} + \alpha_{1} \dot{\phi} - 9 H \dot{\phi}^2 G_{,X} -  3 H \dot{\phi}^{4} G_{,XX}\right) \delta \dot{\phi} +   \left[\frac{k^2}{a^2} \left( \dot{\phi}  G_{,Z} - \dot{\phi}^2  G_{,X} \right) \right]\delta \phi \nonumber \\
    & & + \left( -3 H  \dot{\phi} ^3 F_{,ZZ} - 9 H  \dot{\phi} ^2 F_{,Z} - \alpha_{1}\dot{\phi} ^2 - 6 M_{\rm P}^2 H ^2 + 12 H \dot{\phi}^{3} G_{,X} + 3 H \dot{\phi}^{5} G_{,XX}\right) \Psi  - \frac{k^2}{a^2} 2 k^2 M_{\rm P}^2 \Phi \nonumber \\
    & &  +\left( - 3 \dot{\phi}^2 F_{,Z} + 3 G_{,X}\dot{\phi}^{3} - 6 M_{\rm P}^2 H \right) \dot{\Phi}  = 0 \,. 
\end{eqnarray}
The momentum constraint equation is
\begin{eqnarray}\label{eq:Echi}
   E_{\chi }& \equiv & -\frac{k^2}{a^2} \left( - \dot{\phi}  F_{,Z} + \dot{\phi}^2 G_{,X} \right) \delta \dot{\phi}  + \frac{k^2}{a^2}  \left( \dot{\phi} - 3 H G_{X}  \dot{\phi}^2 -  \ddot{\phi}  F_{,Z} \right) \delta \phi  + \frac{k^2}{a^2} \left(2 M_{\rm P}^2 H + \dot{\phi}^2 F_{,Z} - \dot{\phi}^3 G_{,X} \right) \Psi + \frac{k^2}{a^2} 2 M_{\rm P}^2 \dot{\Phi}  \nonumber \\
   & & + \frac{\theta_{c}}{a}  \left[3 H \dot{\phi}^2 F_{,Z} + \dot{\phi}  \ddot{\phi}  F_{,Z} - \big(\rho_{c} + P_{c} \big) \right] - \sum_{I = b, r}\frac{\theta_{I}}{a}  \left(\rho_{I} + P_{I}\right) = 0 \,.
\end{eqnarray}

The equation of motion for the field $\zeta$ is 
\begin{eqnarray}
    E_{\zeta} & \equiv & \Psi  \left[\frac{k^2}{3 a ^2}+\frac{\alpha \dot{\phi} ^2}{2 M_{\rm P}^{2}}-\frac{ \dot{\phi} ^2 \ddot{\phi}  F_{,ZZ}}{2 M_{\rm P}^{2}}-\frac{3   \dot{\phi}  \ddot{\phi} F_{,Z}}{2 M_{\rm P}^{2}}+\frac{\dot{\phi} ^4 \ddot{\phi}  G_{,XX}}{2 M_{\rm P}^{2}}+\frac{2 \dot{\phi} ^2 \ddot{\phi}  G_{,X}}{M_{\rm P}^{2}}-2 \dot{H} -3 H ^2\right]-\frac{k^2 }{3 a ^2}\Phi - \alpha_{1} \frac{\dot{\phi} }{2 M_{\rm P}^{2}}\delta\dot{\phi}  \nonumber \\
    & & +\sum_{I=r,b,c} \frac{ \delta P_{I} }{2 M_{\rm P}^{2}   }+\delta\ddot{\phi} \left[\frac{   \dot{\phi}F_{,Z}}{2 M_{\rm P}^{2}}-\frac{\dot{\phi} ^2 G_{,X}}{2 M_{\rm P}^{2}}\right] + \frac{  \dot{\phi}  \ddot{\phi}  F_{,ZZ}}{2 M_{\rm P}^{2}}\delta\dot{\phi} +\frac{  \ddot{\phi} F_{,Z}}{2 M_{\rm P}^{2}}\delta\dot{\phi} -\frac{  \dot{\phi} ^3 \ddot{\phi}  G_{,XX}}{2 M_{\rm P}^{2}}\delta\dot{\phi} - \frac{  \dot{\phi}  \ddot{\phi}  G_{,X}}{M_{\rm P}^{2}}\delta\dot{\phi} \nonumber \\
    & & +\dot{\Psi} \left[ -\frac{   \dot{\phi} ^2F_{,Z}}{2 M_{\rm P}^{2}}+\frac{\dot{\phi} ^3 G_{,X}}{2 M_{\rm P}^{2}}-H \right] - 3 H  \dot{\Phi} -\ddot{\Phi}  = 0 \,.
\end{eqnarray}
The shear equation of motion is given by
\begin{equation}
    E_{\rm shear} \equiv M_{\rm P}^2 k^2 (\Psi - \Phi) + \frac{3}{2} a^2 (P_{r} + \rho_{r}) \sigma_{r} = 0 \,. \label{eq:EQshear}
\end{equation}
Finally, the equation of motion for the scalar mode is given by
\begin{eqnarray}
    E_{\delta\phi} & \equiv & \frac{k^2}{a^2} \left[-G_{,XX}\ddot{\phi}  \dot{\phi} ^2 + \alpha_{1} + H  \left(   F_{,Z}-4 G_{,X} \dot{\phi} \right)-2 G_{,X}\ddot{\phi} +   F_{,ZZ}\ddot{\phi} \right] \delta\phi +\delta\ddot{\phi}  \left[\alpha_{1} + 3 H  \left\{   F_{,Z}+\dot{\phi}  \left(-G_{,XX} \dot{\phi} ^2-2 G_{,X}+   F_{,ZZ}\right)\right\}\right] \nonumber \\
    & & +\dot{\Psi}  \dot{\phi} \left[ H  \left\{3 \dot{\phi}  \left(G_{,XX} \dot{\phi} ^2+3 G_{,X}-   F_{,ZZ}\right)-6 F_{,Z}\right\} - \alpha_{1} \right] + 3 \dot{\phi}  \left(G_{,X} \dot{\phi} -  F_{,Z}\right) \ddot{\Phi} + \left[ F_{,Z} \left(3 H  \dot{\phi} +\ddot{\phi} \right)/a \right]\theta_{c}\nonumber \\
    & & -3 \dot{\Phi}  \left[ -G_{,XX}\ddot{\phi}  \dot{\phi} ^3-6 H  G_{,X} \dot{\phi} ^2+\left(\alpha_{1} + 6    H  F_{,Z}-2 G_{,X}\ddot{\phi} +   F_{,ZZ}\ddot{\phi} \right) \dot{\phi} +   F_{,Z}\ddot{\phi} \right] \nonumber \\
    & & -3 \Phi  \left[9 \dot{\phi}  \left(   F_{,Z}-G_{,X} \dot{\phi} \right) H ^2+3 \left\{ -G_{,XX}\ddot{\phi}  \dot{\phi} ^3+\left(\alpha_{1} - 2 G_{,X}\ddot{\phi} +   F_{,ZZ}\ddot{\phi} \right) \dot{\phi} +  F_{,Z}\ddot{\phi} \right\} H -3 G_{,X} \dot{H}  \dot{\phi} ^2+3    F_{,Z} \dot{H}  \dot{\phi} +\alpha_{1}\ddot{\phi} \right] \nonumber \\
    & & +3 \delta \dot{\phi}  \left[3 \left\{   F_{,Z}+\dot{\phi}  \left(-G_{,XX} \dot{\phi} ^2-2 G_{,X}+   F_{,ZZ}\right)\right\} H ^2+\left(-\ddot{\phi} G_{,XXX} \dot{\phi} ^4-5 G_{,XX}\ddot{\phi}  \dot{\phi} ^2+  \ddot{\phi}  F_{,ZZZ} \dot{\phi} \right. \right. \nonumber \\
    & & \left. \left. +\alpha_{1}-2 G_{,X}\ddot{\phi} +2    F_{,ZZ}\ddot{\phi} \right) H +\dot{H}  \left\{  F_{,Z}+\dot{\phi}  \left(-G_{,XX} \dot{\phi} ^2-2 G_{,X}+   F_{,ZZ}\right)\right\}\right] \nonumber \\
    & & +\Psi  \left[3 \dot{H}  G_{,XX} \dot{\phi} ^4+9 G_{,X} \dot{H}  \dot{\phi} ^2-3    \dot{H}  F_{,ZZ} \dot{\phi} ^2-6    F_{,Z} \dot{H}  \dot{\phi} +k^2/a^2 \left(   F_{,Z}-G_{,X} \dot{\phi} \right) \dot{\phi} \right. \nonumber  \\
    & & -9 H ^2 \left\{2    F_{,Z}+\dot{\phi}  \left(-G_{,XX} \dot{\phi} ^2-3 G_{,X}+   F_{,ZZ}\right)\right\} \dot{\phi} -\alpha_{1}\ddot{\phi} -3 H  \left\{-\ddot{\phi}  G_{,XXX} \dot{\phi} ^5-7 G_{,XX}\ddot{\phi}  \dot{\phi} ^3+  \ddot{\phi}  F_{,ZZZ} \dot{\phi} ^2 \right. \nonumber \\ 
    & & \left. \left. +\left(\alpha-6 G_{,X}\ddot{\phi} +4    F_{,ZZ}\ddot{\phi} \right) \dot{\phi} +2    F_{,Z}\ddot{\phi} \right\}\right] = 0 \,.
\end{eqnarray}

\section{Quasi-Static Approximation and Effective Gravitational constants} \label{sec:QSA_limit}
In this section, we study the evolution of the baryonic and cold dark matter over-densities using the quasi-static approximation (QSA) limit. Let us start from the matter equation 
\begin{eqnarray}
    E_{\theta_{b}} & \equiv &  \dot{\delta}_{b}  + \frac{\theta_{b}}{a} = 0 \, , \label{eq:EthetamQSA}\\
    E_{\delta_{b}} & \equiv & \dot{\theta}_{b} + H \theta_{b}  - \frac{k^2}{a} \Psi = 0 \, \label{eq:EdeltamQSA}.
\end{eqnarray}
where we have considered that the contribution of matter is for the baryon, i.e. $P_{b}=0$. Then we have 
\begin{equation}\label{eq:EQdeltam2nd}
    \ddot{\delta}_{b} + 2H \dot{\delta}_{b} + \frac{k^2}{a^2} \Psi = 0  \, .
\end{equation}

Following the same procedure for the dark matter sector we have 
\begin{eqnarray}
    E_{\theta_{c}} & \equiv &  \dot{\delta}_{c}  + \frac{\theta_{c}}{a} = 0 \, , \label{eq:EQthetadmQSA}
\end{eqnarray}
and the equation of motion for the velocity potential, Eq.~(\ref{EQ:EQdeltadm}) remains the same in the quasi-static approximation limit. Taking the time derivative of the Eq.~(\ref{eq:EQthetadmQSA}) and substituting the $\dot{\theta}_{c}$ and $\theta_{c}$, with Eq.~(\ref{EQ:EQdeltadm}), resulting in
\begin{equation}
    E_{\delta} (\ddot{\delta}_{c}, \dot{\delta}_{c}, \Phi, \delta \dot{\phi}, \delta \phi)\,,
\end{equation}
where, we have used $\Phi = \Psi$ from the shear equation Eq.~(\ref{eq:EQshear}), since $\sigma = 0$ for baryons. 

Now we need to find the solution for $\Phi$ and $\delta \phi$ in the QSA limit. On taking the QSA limit of the Eq.~(\ref{EQ:EQalpha}), we get
\begin{equation}\label{eq:E_alpha_QSA}
    E_{\alpha}^{\rm QSA} \equiv - \rho_{b} \delta_{b} - \rho_{c} \delta_{c} + \frac{k^2}{a^2}\left[ -2 M_{\rm P}^{2} \Phi + \delta \phi \left(\dot{\phi}F_{3,Z} + \dot{\phi}^2G_{,X} \right) \right] \simeq 0 \,,
\end{equation}
Notice that in the quasi-static approximation, the momentum equation remains the same. Now taking the time derivative of the Eq.~(\ref{eq:E_alpha_QSA}), and from the linear combination of Eq.~(\ref{eq:Echi}), and using Eq.~(\ref{eq:E_alpha_QSA}), we arrive at 
\begin{eqnarray}\label{EQ:EQdphiQSA}
    \frac{k^2}{a^2}\, A\, \delta\phi  & \simeq & 2  M_{\rm P}^{2} \dot{\delta}_{c}  F_{,Z} \left(3 H  \dot{\phi} +\ddot{\phi} \right)+\delta_{c}  \rho_{c}  \dot{\phi}  \left(    F_{,Z}-\dot{\phi}  G_{,X}\right)+\delta_{b}  \rho_{b}  \dot{\phi}  \left( F_{,Z}-\dot{\phi}  G_{,X}\right) \,,
\end{eqnarray}
where  $A$ is given by 
\begin{eqnarray}
    A & \equiv & \alpha_{1} 2 M_{\rm P}^{2}+2   M_{\rm P}^{2} \ddot{\phi}  F_{,ZZ}-2   \dot{\phi} ^3 F_{,Z} G_{,X}+  \dot{\phi} ^2 F_{,Z}^2+2  M_{\rm P}^{2} H  F_{,Z}-2 M_{\rm P}^{2} \dot{\phi} ^2 \ddot{\phi} G_{,XX}-8 M_{\rm P}^{2} H  \dot{\phi}  G_{,X} \nonumber \\
    & & -4 M_{\rm P}^{2} \ddot{\phi}  G_{,X}+\dot{\phi} ^4 G_{,X}^2 \,.
\end{eqnarray}
Now we have solution for the $\delta \phi$, given by Eq.~(\ref{EQ:EQdphiQSA}) $\Psi$ (by solving Eq.~(\ref{eq:E_alpha_QSA}), and substituting $\delta \phi$).

On inserting the solution for $\Psi$ and $\delta \phi$ we can write the second-order equation for the baryonic over-density and the cold dark matter over-density, we get
\begin{equation}
    \ddot{\delta}_{b} + 2 H \dot{\delta}_{b} + \Gamma_{bc} \dot{\delta}_{c}- \frac{3}{2} \frac{H^2}{G} a^2 (G_{ bb}\Omega_{b} \delta_{b} + G_{bc}\Omega_{c} \delta_{\rm c} ) = 0\,, \label{eq:delta_b}
\end{equation}
 and 
 \begin{eqnarray}
\label{eq:delta_c2ndorder}
    \ddot{\delta}_{ c} +2 H \Gamma_{cc} \, \dot{\delta}_{ c} + \Gamma_{cb} \dot{\delta}_{b} - \frac{3}{2} \frac{H^2}{G} a^2 (G_{ cb}\Omega_{ b} \delta_{ b} + G_{cc}\Omega_{ c} \delta_{ c} ) = 0\,, \label{eq:delta_c}
\end{eqnarray}
where
\begin{equation}
    \Gamma_{bc} \equiv \frac{  \dot{\phi}  F_{,Z} \left(3 H  \dot{\phi} +\ddot{\phi} \right) \left( F_{,Z}-\dot{\phi} G_{,X}\right)}{A} \,,
\end{equation}
and
\begin{equation}
    G_{bb} = G_{bc} = \frac{2 M_{\rm P}^{2} \left[\alpha_{1}+\gamma  \ddot{\phi}  F_{,ZZ}+H  \left(\gamma  F_{,Z}-4 \dot{\phi}  G_{,X}\right)+\dot{\phi} ^2 \ddot{\phi}  \left(-G_{,XX}\right)-2 \ddot{\phi}  G_{,X}\right]}{A} \,.
\end{equation}

The expression for $\Gamma_{cc}$, $\Gamma_{cb}$, $G_{cb}$ and $G_{cc}$ is given in Appendix \ref{apdx:Gammas_and_G}.

\section{Concrete Model}
\label{sec:concrete_model}
Until now, we considered the model to be completely general. We have already found the background equations of motion, no-ghost condition, Laplace instability conditions, and the perturbed equations of motion. Now we introduce a concrete model
\begin{equation}
    G(X) = \beta\frac{X}{M^3}\,, \qquad F(Z)  = \gamma \frac{Z^{2}}{M^{3}}\,. 
    \label{eq:model}
\end{equation}
This model has a tracking behaviour $\dot{\phi}H = \text{constant}$~\cite{DeFelice:2010pv}. In the following subsections, we study the background dynamics, no-ghost conditions and evolution of the matter over-densities with the concrete model.

\subsection{Background and Ghost conditions}
\subsubsection{Background dynamics}
We introduce the following redefinition, which has been introduced in~\cite{DeFelice:2010pv}
\begin{equation}
    r_{1} \equiv \frac{H_{\rm dS} \dot{\phi}_{\rm dS}}{H \dot{\phi}}\,, \qquad r_{2} \equiv \left( \frac{\dot{\phi}}{\dot{\phi}_{\rm dS}} \right)^{4} \frac{1}{r_{1}} \,,
\end{equation}
where $H_{\rm dS}$ and $\dot{\phi}_{\rm dS}$ is the value of Hubble expansion and the value of $\dot{\phi}$ at the de Sitter solution respectively. Since at de Sitter, all the matter content will be absent, we can keep $\gamma = 0$ at the de Sitter.

Now we can find that 
\begin{equation}
    \frac{r_{1}'}{r_{1}} = - \frac{\dot{H}}{H^2{}} - \frac{\ddot{\phi}}{H \dot{\phi}}\, , \qquad \text{and} \qquad \frac{r_{2}'}{r_{2}} = - \frac{r_{1}'}{r_{1}} - \frac{4 \ddot{\phi}}{H \dot{\phi}} \,,
\end{equation}
where the prime denotes the derivative with respect to the variable $\mathcal{N} = \ln (a)$ and over dot denotes the derivative with respect to time. We can use Eq. (\ref{eq:2ndFE}) and Eq. (\ref{eq:eqdphiBG}) to substitute $\ddot{\phi}$ and $\dot{H}$. 

Now considering the Eq.~(\ref{eq:1stFE}) and Eq.~(\ref{eq:2ndFE}) at de Sitter, i.e. all the matter content vanishing including the interaction term and $\ddot{\phi} = 0$, $\dot{H} = 0$. Then we can find that 
\begin{equation}
    \alpha_{1} = \frac{6}{x_{\rm dS}^{2}}\,, \qquad \text{and} \qquad \beta = \frac{2}{x_{\rm dS}}, \qquad \text{where} \qquad x_{\rm dS} = \frac{\dot{\phi}_{\rm dS}}{M_{\rm P} H_{\rm dS}}\,.
\end{equation}

We also introduce the density variables
\begin{equation}
    \Omega_{i} \equiv \frac{\rho_{i}}{3 M_{\rm P}^{2} H^{2}} \,.
\end{equation}
We can write dark energy energy density as
\begin{equation}\label{eq:OmegDEinr1r2}
    \Omega_{\rm DE} = - \frac{1}{6} \alpha_{1} x_{\rm dS}^{2} r_{1}^{3} r_{2} - \beta x_{\rm dS}^3 r_{1}^{2} r_{2} - 2 \gamma x_{\rm dS} r_{1}^2 r_{2} \,,
\end{equation}
where we have also used the relation 
\begin{equation}
    M^{3} \equiv H_{\rm dS}^{2} M_{\rm P}\,.
\end{equation}
We also use the Friedmann constrain to replace $\Omega_{b}$
\begin{equation}
    \Omega_{b} = 1 - \Omega_{r} - \Omega_{c} - \Omega_{\rm DE} \,.
\end{equation}

After choosing the concrete model and making the field redefinitions to new variables, we can rewire the background equations of motion in terms of new variables. The evolution of the variable $r_{1}$ is given by
\begin{equation}\label{eq:EQr1}
    \frac{r_{1}' }{r_{1} }-\frac{\left(r_{1} +\gamma  x_{\rm dS}^3-1\right) \left(6 \left(\gamma x_{\rm dS}^3-1\right) r_{1} ^2 r_{2} +3 r_{1} ^3 r_{2} - \Omega_{r}-9\right)}{2 \left(\left(\gamma  x_{\rm dS}^3-1\right)^2 r_{1} ^2 r_{2} -r_{1} -2 \gamma  x_{\rm dS}^3+2\right)} = 0 \,, 
\end{equation}
the evolution of $r_{2}$ is given by
\begin{equation}\label{eq:EQr2}
    \frac{r_{1}' }{r_{1} }+\frac{r_{2}' }{r_{2} } -\frac{2 \left(3 \left(\gamma  x_{\rm dS}^3-1\right) r_{1} ^3 r_{2} +6 r_{1} +\left(1-\gamma x_{\rm dS}^3\right) \Omega_{r}+3 \gamma  x_{\rm dS}^3-3\right)}{\left(\gamma  x_{\rm dS}^3-1\right)^2 r_{1} ^2 r_{2} -r_{1} -2 \gamma  x_{\rm dS}^3+2} = 0 \,.
\end{equation}
The evolution of the variable $\Omega_{r}$ is,
\begin{equation}\label{eq:EQOmegar}
    4 \frac{\Omega_{r}' }{\Omega_{r} } - 10 \frac{r_{1}' }{r_{1}} - 2 \frac{ r_{2}' }{r_{2} }+16 = 0 \,.
\end{equation}
Finally, we can also find the differential equation for $\Omega_{c}$ as
\begin{equation} \label{eq:EQOmegac}
    \frac{\Omega_{c}' }{\Omega_{c}}-\frac{5 r_{1}' }{2 r_{1} }-\frac{r_{2}' }{2 r_{2} }+3 = 0 \,.
\end{equation}

\begin{figure}[h]
    \centering
    \includegraphics[scale=1.2]{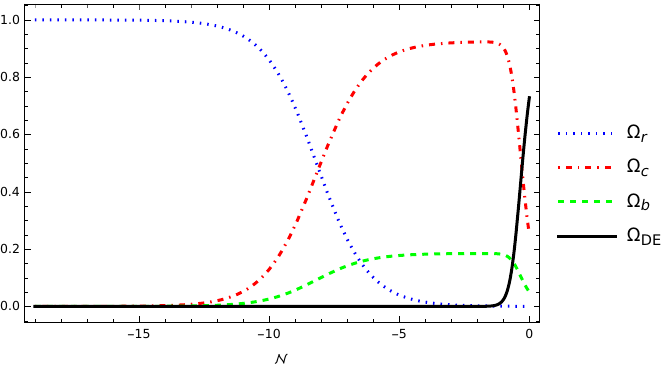}
    \caption{Evolution of $\Omega_{\rm DE}$, $\Omega_{r}$, $\Omega_{c}$, $\Omega_{b}$. This plot is generated by solving the coupled background equations for $r_{1}$, $r_{2}$, and $\Omega_{r}$ as given in Eq.~(\ref{eq:EQr1}), Eq.~(\ref{eq:EQr2}), Eq.~(\ref{eq:EQOmegar}). The differential equation for $\Omega_{c}$ is given in Eq.~(\ref{eq:EQOmegac}). $\Omega_{\rm DE}$ is given in terms of the variable $r_{1}$ and $r_{2}$ in Eq.~(\ref{eq:OmegDEinr1r2}). We also have chosen $\gamma = 0.01$.}
    \label{fig:Omega_i_evo}
\end{figure}

We solve the coupled differential equations for the variables $r_{1}$, $r_{2}$, $\Omega_{r}$, and $\Omega_{c}$ numerically giving the following initial conditions $r_{1} = 1$, $r_{2} = 1.2 \times 10^{-58}$, $\Omega_{r} = 0.99998$, $\Omega_{c} = 1.85 \times 10^{-5}$ and $\Omega_{b} = 3.7 \times 10^{-6}$ at $\mathcal{N} = \ln(a) = -19$. The resulting evolution is presented in the figure~\ref{fig:Omega_i_evo}. 

The figure~\ref{fig:Omega_i_evo} shows that, in this model, the behaviour of the composition of our universe at the background level behaves as expected. The new contribution in this model is the interaction $Z$ in the $G_{3}$ term in the Lagrangian, which is proportional to $\gamma$, where we choose $\gamma = 0.01$. 

\subsubsection{Stability conditions}
In this part, we show the behaviour of the no-ghost condition and the square of the speed of propagation for the scalar mode. After choosing the concrete model and making the field redefinition, the no-ghost condition for the scalar mode  becomes
\begin{equation}
    Q_{s}/M_{\rm P}^{2} = \frac{3 r_{1} ^2 r_{2}  (r_{1} +2 \gamma_{x_{\rm dS}}-2) \left[(\gamma_{x_{\rm dS}}-1)^2 r_{2} (r_{1} +2 \gamma_{x_{\rm dS}}-2)-1\right]}{\left[2 (\gamma_{x_{\rm dS}}-1)^2 r_{1}  r_{2} +(\gamma_{x_{\rm dS}}-1) r_{1} ^2 r_{2} -1\right]^2} > 0 \,,
\end{equation}
where $\gamma_{x_{\rm dS}} \equiv \gamma x_{\rm dS}^{3}$.

\begin{figure}[h]
    \centering
    \includegraphics[scale=1.2]{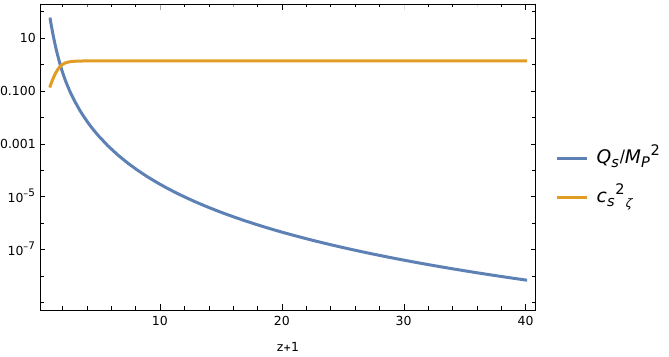}
    \caption{Behaviour of the $Q_{s}/M_{\rm P}^{2}$ and $c_{s}^{2}{}_{\zeta}$. We have used the same initial conditions to solve the background differential equations with $\gamma = 0.01$, and $x_{\rm dS} = 1$.}
    \label{fig:ghost}
\end{figure}

The behaviours of the $Q_{s}/M_{\rm P}^{2}$, and the $c_{s}^{2}{}_{\zeta}$ is given in the figure~\ref{fig:ghost}. We have used same initial values for the variables $r_{1}$, $r_{2}$, $\Omega_{r}$, and $\Omega_{c}$ as for the figure~\ref{fig:Omega_i_evo}. 

\subsection{Dynamics of matter over-densities}
\begin{figure}[h]
    \centering
    \includegraphics[scale=1.2]{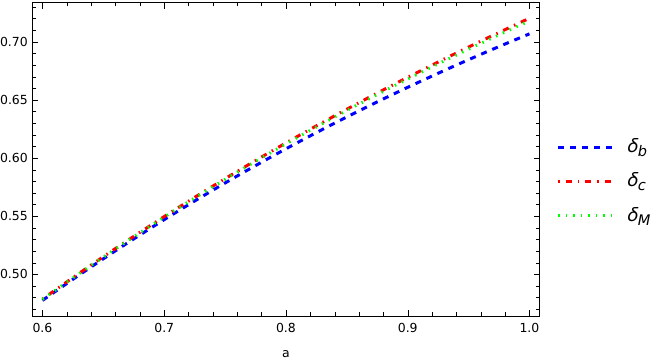}
    \caption{Evolution of the fields $\delta_{c}$ (dot dashed red), $\delta_{b}$ (dashed blue), and the total matter over-density $\delta_{M}$ (dotted red).}
    \label{fig:delta_evo}
\end{figure}

Now we study the evolution of the baryonic and cold dark matter over-densities. We solve numerically the Eq. (\ref{eq:delta_b}) and Eq. (\ref{eq:delta_c}), using the concrete model Eq. (\ref{eq:model}) we have introduced in this section. The evolution is plotted in the figure \ref{fig:delta_evo}. We used the initial condition $\delta'_{m = c, b} = \delta_{m = c, b} = 0.017$ at the redshift $z = 39.45$. We also plot the total matter over-density given by
\begin{equation}
    \delta_{\rm M} = \frac{\Omega_{b} \delta_{b}}{(\Omega_{b} + \Omega_{c})} + \frac{\Omega_{c} \delta_{c}}{(\Omega_{b} + \Omega_{c})} \,. \label{eq:deltaM}
\end{equation}
We can clearly see in the figure~\ref{fig:delta_evo} that there is an enhanced growth to the cold dark matter component, with respect to the baryonic component, hence the total matter component also shows an enhanced growth. We will see in the next section the effect of this modification in the redshift space distortion observable.

\section{Testing the model using genetic algorithm} \label{sec:GA}
Until now, we have computed the background equations of motion, ghost conditions, perturbation equations as well as the equations of motion for the baryon and the cold dark matter over-densities. Also, we analysed their evolution along the redshift after introducing a concrete model for the free functions. In this section, we move on to use the genetic algorithm to make an initial test of the theory we introduced. 

\subsection{$H(z)$ data}
\begin{figure}[h]
    \centering
    \includegraphics[scale=1.2]{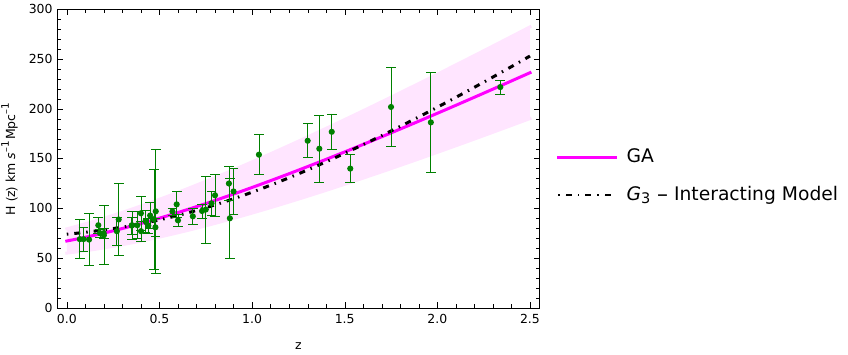}
    \caption{This plot shows the $H(z)$ curve predicted by the genetic algorithm (magenta line), using the data points given in the figure (green points) with error bars. The light shaded region is the $1 \sigma$ error predicted by the algorithm. The dotted dashed line shows the $H(z)$ given by the $G_{3}$ interacting model.}
    \label{fig:H_of_z_GA}
\end{figure}

In this subsection, we study the compatibility of background dynamics with genetic algorithm, using the cosmological data. In particular, we use $H(z)$ data compilations used in~\cite{Alestas:2022gcg}. These data are given in table~\ref{tab:H_of_z_data}, which consists of 39 data points. Now we use the genetic algorithm code provided in the~\cite{Alestas:2022gcg}, to predict the $H(z)$ curve using the data provided. We use the set of orthogonal functions, also known as grammar, to be $x$ and $x^{x}$. We have chosen the crossover rate to be $0.75$ and the mutation rate to be $0.25$, with 1500 generations. If the $\chi^2$ is not improved for a few hundred generations the algorithm is stopped. The resulting curve is plotted in the figure~\ref{fig:H_of_z_GA}.

The function $H(z)$ predicted by the genetic algorithm is 
\begin{equation}
    H(z)^{GA} = H_{\rm marg}^{GA} \sqrt{z (0.453104 z+1.01544)^2+1}
\end{equation}
with the $\chi_{\rm min}^{2} = 1066.87$, and $H_{\rm marg}^{GA}$ is the marginalised value of the Hubble function today $H_{\rm marg}^{GA} = 67.54$. 

A methodology of error estimation in the genetic algorithm is developed in~\cite{Nesseris:2012tt} using the path integral technique. We use the same algorithm to estimate the error up to $1 \sigma$ which is shown in the figure. We also over plot the $H(z)$ function from the background dynamics of the model we have considered, and we choose $H_{\rm dS} = 67$. It is clear from the figure~\ref{fig:H_of_z_GA} that, $H(z)$ function is well inside the $1 \sigma$ region. In other words, the genetic algorithm validates the model at the level of background dynamics.

\begin{table}
    \centering
    \begin{tabular}{| c | c | c | c | c|}
    \hline
    Redshift ($z$)   &   $H(z)$  &   $1 \sigma$  & Ref. \\
     \toprule    
    0.09    &	69    &	  12  &   \cite{Stern:2009ep} \\
    0.17    &	83    &   8   &   \cite{Stern:2009ep} \\
    0.27    &	77    &   14   &   \cite{Stern:2009ep} \\
    0.40    &   95    &   17   &   \cite{Stern:2009ep} \\
    0.48    &	97    &   62   &   \cite{Stern:2009ep} \\
    0.88    &	90    &   40   &   \cite{Stern:2009ep} \\
    0.90    &	117   &   23   &   \cite{Stern:2009ep} \\
    1.30    &	168   &   17   &   \cite{Stern:2009ep} \\
    1.43    &	177   &   18   &   \cite{Stern:2009ep} \\
    1.53    &	140   &   14   &   \cite{Stern:2009ep} \\
    1.75    &	202   &   40   &   \cite{Stern:2009ep} \\
    0.44    &	82.6  &   7.8   &   \cite{Blake:2012pj} \\
    0.60    &	87.9  &   6.1   &   \cite{Blake:2012pj} \\
    0.73    &	97.3  &   7.0   &   \cite{Blake:2012pj} \\
    0.179   &	75    &   4   &   \cite{Moresco:2012jh} \\
    0.199   &	75    &   5   &   \cite{Moresco:2012jh} \\
    0.352   &	83    &   14   &   \cite{Moresco:2012jh} \\
    0.593   &	104   &   13   &   \cite{Moresco:2012jh} \\
    0.68    &	92    &   8   &   \cite{Moresco:2012jh} \\
    0.781   &	105   &   12   &   \cite{Moresco:2012jh} \\
    0.875   &	125   &   17   &   \cite{Moresco:2012jh} \\
    1.037   &	154   &   20   &   \cite{Moresco:2012jh} \\
    0.35    &	82.7  &   8.4   &   \cite{Chuang:2012qt} \\
    0.07    &	69.0  &   19.6   &   \cite{Zhang:2012mp} \\
    0.12    &	68.6  &   26.2   &   \cite{Zhang:2012mp} \\
    0.20    &	72.9  &   29.6   &   \cite{Zhang:2012mp} \\
    0.28    &	88.8  &   36.6   &   \cite{Zhang:2012mp} \\
    0.57    &	96.8  &   3.4   &   \cite{BOSS:2013rlg} \\
    2.34    &	222.0 &   7.0   &   \cite{BOSS:2014hwf} \\
    1.363   &	160.0 &   33.6   &   \cite{Moresco:2015cya} \\
    1.965   &	186.5 &   50.4   &   \cite{Moresco:2015cya} \\
    0.3802  &	83.0  &   13.5   &   \cite{Moresco:2016mzx} \\
    0.4004  &	77.0  &   10.2   &   \cite{Moresco:2016mzx} \\
    0.4247  &	87.1  &   11.2   &   \cite{Moresco:2016mzx} \\
    0.4497  &	92.8  &   12.9   &   \cite{Moresco:2016mzx} \\
    0.4783  &	80.9  &   9.0   &   \cite{Moresco:2016mzx} \\
    0.47    &	89    &   50   &   \cite{Ratsimbazafy:2017vga} \\
    0.75    &	98.8  &   33.6   &   \cite{Borghi:2021rft} \\
    0.80    &	113.1 &   20.73   &   \cite{Jiao:2022aep} \\
    \hline
    \end{tabular}
    \caption{This table shows the $H(z)$ data compilations and the corresponding $1 \sigma$  error bars from various observations. We have also given the corresponding references.}
    \label{tab:H_of_z_data}
\end{table}

\subsection{Redshift Space Distortion data}
We have seen that the model we have introduced can modify the effect of gravitational coupling and hence can modify the evolution of the baryon and cold dark matter over-densities. We solved numerically the evolution of these over-densities using the concrete model we have introduced in the previous section. After solving the equations, we can compute $f \sigma_{8}$ for this model. Then we can use a genetic algorithm to test if this model is viable at the linear perturbation level. We will find that, since this model shows an enhanced behaviour in the growth of the structures the value of $f \sigma_{8}$ in the redshift close to today lies outside the $1\sigma$ error bar determined by the genetic algorithm. Even though, this model shows less preference with the genetic algorithm, we have not tried complete parameter estimation to find the value of the new parameter $\gamma$ corresponding to the new interaction we have introduced.

The redshift space distortion observable is given by
\begin{equation}
    f \sigma_{8} = f(a) \sigma(a) = \frac{\sigma_{8,0}}{\delta_{m,0}} a \frac{d \delta_{m}}{d a} \,,
\end{equation}
where 
\begin{equation}
    f(a) = \frac{d \ln(\delta_{m})}{d a} \,, \qquad \sigma(a) = \sigma_{8,0} \frac{\delta_{m}}{\delta_{m,0}}\,.
\end{equation}
$\delta_{m,0}$ is the value of $\delta_{m}$ today. The $\delta_{m}$ we consider from this model is for the total matter content given as in Eq.~(\ref{eq:deltaM}), and we choose $\sigma_{8, 0} = 0.8$ and $\delta_{m,0} = 0.718064$.

We consider the redshift space distortion data that as been complied in the~\cite{Alestas:2022gcg}, the data is given in the table \ref{table:growth_data}. We use these data as the input to the genetic algorithm to predict  $f\sigma_{8}$ curve model independently~\cite{Arjona:2020kco, Alestas:2022gcg}. We use the same error estimation method developed in~\cite{Nesseris:2012tt} to estimate the error up to $1 \sigma$.
\begin{table}
    \centering
    \begin{tabular}{| c | c | c | c | c| }
    \hline
          Redshift ($z$) & $f\sigma_{8}$ & $1\sigma$ & $\Omega_{m,0}$ & Ref. \\
          \toprule
          0.17 & 0.510 & 0.060 & 0.3 & \cite{Song:2008qt}  \\
          0.02 & 0.314 & 0.048 & 0.266 & \cite{Davis:2010sw, Hudson:2012gt}  \\
          0.02 & 0.398 & 0.065 & 0.3 & \cite{Hudson:2012gt, Turnbull:2011ty}  \\
          0.44 & 0.413 & 0.080 & 0.27 & \cite{Blake:2012pj}  \\
          0.60 & 0.390 & 0.063 & 0.27 & \cite{Blake:2012pj}  \\
          0.73 & 0.437 & 0.072 & 0.27 & \cite{Blake:2012pj}  \\
          0.18 & 0.360 & 0.090 & 0.27 & \cite{Blake:2013nif}  \\
          0.38 & 0.440 & 0.060 & 0.27 & \cite{Blake:2013nif}  \\
          1.40 & 0.482 & 0.116 & 0.27 & \cite{Okumura:2015lvp}  \\
          0.02 & 0.428 & 0.0465 & 0.3 & \cite{Huterer:2016uyq}  \\
          0.60 & 0.550 & 0.120 & 0.3 & \cite{Pezzotta:2016gbo}  \\
          0.86 & 0.400 & 0.110 & 0.3 & \cite{Pezzotta:2016gbo}  \\
          0.03 & 0.404 & 0.0815 & 0.312 & \cite{Qin:2019axr}  \\
          0.013 & 0.46 & 0.060 & 0.315 & \cite{Avila:2021dqv} \\
          0.15 & 0.530 & 0.160 & 0.31 & \cite{eBOSS:2020yzd} \\
          0.38 & 0.500 & 0.047 & 0.31 & \cite{eBOSS:2020yzd} \\
          0.51 & 0.455 & 0.039 & 0.31 & \cite{eBOSS:2020yzd} \\
          0.70 & 0.448 & 0.043 & 0.31 & \cite{eBOSS:2020yzd} \\
          0.85 & 0.315 & 0.095 & 0.31 & \cite{eBOSS:2020yzd} \\
          1.48 & 0.462 & 0.045 & 0.31 & \cite{eBOSS:2020yzd} \\
          \hline
    \end{tabular}
    \caption{This table shows the value of $f \sigma_{8}$, corresponding $1 \sigma$ error bar, which is compiled in~\cite{Alestas:2022gcg}. The table also gives the corresponding value of $\Omega_{m,0}$ used, which is the assumption of the fiducial background.}
    \label{table:growth_data}
\end{table}

\begin{figure}[h]
    \centering
    \includegraphics[]{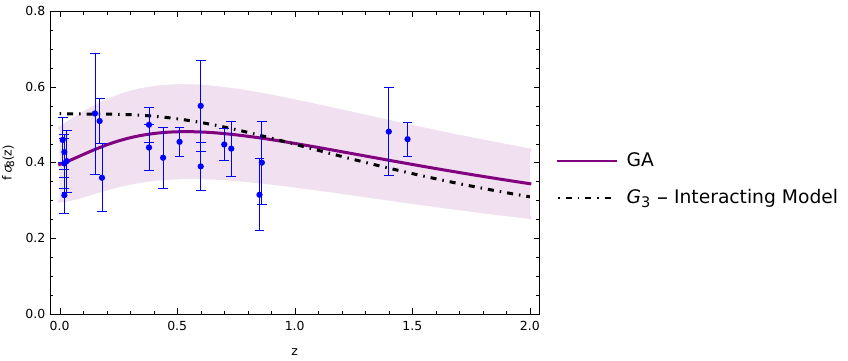}
    \caption{This plot shows the function generated by the genetic algorithm Eq.~(\ref{eq:fsigma8_GA}), the purple line. The purple shaded region is the $1 \sigma$ error bar generated. The dotted-dashed black line is the $f\sigma_{8}$ curve predicted by the theory.}
    \label{fig:fsigma8GA}
\end{figure}

The set of orthogonal functions (grammar) that we have used in our GA analysis are $x$ and $x^{x}$. We have chosen the crossover rate and the mutation rate as $0.75$ and $0.25$ respectively, with $1500$ generations. The algorithm will be terminated once the $\chi^{2}$ does not improve for a few hundred generations. The resulting curve is given in the figure~\ref{fig:fsigma8GA}. We also plotted the $f\sigma_{8}$ predicted by the model choosing one particular value of $\gamma= 0.01$, which is the new parameter in the model. It is clear that the value of $f\sigma_{8}$ in the redshift close to today deviates from the $1\sigma$ predicted by GA, using the redshift space distortion (RSD) data compilation as given in the table \ref{table:growth_data}.

The best function concluded by the GA with the given RSD data is 
\begin{equation}\label{eq:fsigma8_GA}
    f \sigma_{8}^{(GA)} (a) = a - a^4 \left(-0.988584 + 0.206888 a^{2.90671} + 0.0165351 a^{4.5531} - 0.122587 \big(0.417286^{0.417286 a} \times a^{0.417286 a} \big)^{5.97559} \right)^{2}\,,
\end{equation}
with the $\chi^2_{\rm min} = -17.173$.

It is true that from the GA analysis, the model is less preferred with the RSD data set. This enhancement in the growth can be understood from the evolution of the matter over-densities, since there is an enhanced growth of the over-densities. This is due to the fact that the ratio of effective gravitational coupling to $G_{N}$ grows above unity in the dark sector domination as shown in the appendix \ref{apdx:geff}.

In fact, we have only analysed this model with a particular value of the new parameter. It would be interesting to know if there are other values for the $\gamma$, which can fit with the data. To study this we need to do a complete parameter estimation using Markov chain Monte Carlo (MCMC) methods. This will also allow us to confront the model with other data sets like cosmic microwave background and other large scale structure data, which will be done in a future work.

\section{Conclusion}\label{sec:conclusion}
In this manuscript, we have attempted to develop a new interacting scalar dark energy, with only momentum transfer, by adding the term $G_{3}(X, Z) \nabla_{\mu} \nabla^{\mu} \phi = G(X)\nabla_{\mu} \nabla^{\mu} \phi + F(Z) \nabla_{\mu} \nabla^{\mu} \phi$, where $Z = u^{\mu}\nabla_{\mu}\phi$. We started the work by introducing the model in the section~\ref{sec:thoery}. In the literature, the momentum transfer interacting model is usually introduced by adding the term $f(Z)$ to the Lagrangian. However, we know that the most general scalar tensor theory is described by Horndeski theory and degenerate higher order scalar tensor theory. When including the interaction with dark matter, there is no restriction to have a term like $F(Z) \nabla_{\mu} \nabla^{\mu} \phi $. 

Then in section~\ref{sec:background_eom} assuming the homogeneous and isotropic flat FLRW metric, we find the background equations of motion. We assume that the theory is shift symmetric, and then the expansion is driven by the kinetic term. Then we proceed to study the ghost conditions for the propagating modes, in section \ref{sec:ghost_condition}. We have two standard tensor modes and one additional scalar mode, which is responsible for the accelerated expansion of the universe, other than the matter scalar modes arising from the matter action. We find the no-ghost condition and the Laplace instability conditions for this model. 

In section~\ref{sec:perturbation_eom}, we find the linear perturbation equations of motion in Newtonian gauge. We can see from the dark matter equations of motion that the continuity equation is not modified, rather the Euler equation is modified. This indicates the interaction is only a momentum transfer. Using the linear perturbation equation, we find the evolution of the baryonic and cold dark matter over-densities in the QSA limit in section~\ref{sec:QSA_limit}. We found that the gravitational couplings to the baryonic and cold dark matter over-densities are modified. 

After deriving the analytic calculation, we introduce a concrete model in the section~\ref{sec:concrete_model}. This model behaves as a tracking solution, that is $\dot{\phi} H = \text{constant}$. Using this concrete model after introducing a field redefinitions, we solve the background equations of motion numerically, and plot the behaviours of the background quantities like $\Omega_{b}$, $\Omega_{c}$, $\Omega_{r}$ and $\Omega_{\rm DE}$, which is shown in the figure~\ref{fig:Omega_i_evo}. Then we study the behaviour of the no-ghost and Laplacian instability conditions with the concrete model along the evolution of the universe, figure~\ref{fig:ghost}. 

Then we solved the equations of motion for the over-densities of baryon and the cold dark matter numerically. The resulting behaviour is shown in the figure~\ref{fig:delta_evo}. In section~\ref{sec:GA} we move to test the model with the $H(z)$ data and the RSD data ($f\sigma_{8}$) given in the table~\ref{tab:H_of_z_data} and~\ref{table:growth_data}, respectively using the genetic algorithm. The resulting figure of background test $H(z)$ data is given in the figure~\ref{fig:H_of_z_GA}. In the figure~\ref{fig:H_of_z_GA}, it is clear that $H(z)$ function of the model we have introduced is well inside the $1\sigma$ region of the error bar predicted by the genetic algorithm. For the perturbation, that is for the RSD data, the resulting figure is shown in~\ref{fig:fsigma8GA}. We found that towards the redshift close to today the value of $f\sigma_{8}$ lies outside the $1\sigma$ region inferred by the genetic algorithm pipeline using the data provided. This enhanced growth is due to the fact that the ratio of the effective gravitational coupling to the $G_{N}$ grows towards the dark energy domination era as shown in the figure~\ref{fig:Gcc_Gbb}. Even though, we find a less preference of this model with $f\sigma_{8}$ data, we have only tested the model with only one value to the new parameter $\gamma$, corresponds to the new interaction we have introduced. To completely rule out this model (at the level of background and the linear perturbation), we need to confront the model with other cosmological data that are currently available from cosmic microwave background radiation and the large scale structures. In other words, we need to do an MCMC simulation and make the parameter estimation with this model.

There are several future directions that are interesting to pursue, gaining intuitions from the current work. One direction is to generalize the concrete model to a general tracker solution with $\dot{\phi} H^{p} = \text{constant}$ as explored in~\cite{DeFelice:2011bh}, and following similar analysis and see if the new parameter $p$ can help in reducing the $f\sigma_{8}$ in to $1\sigma$ region of the genetic algorithm pipeline, using the same data. Another direction is to generalise the interacting model itself as studied in the context of Horndeski theory~\cite{DeFelice:2010nf}. Then it would be interesting to see the behavior of the $f \sigma_{8}$. Even though, we did not see any additional propagating modes on the perturbation around the FLRW background, it would be interesting to see if any spurious modes are present in this model. One approach is to do a perturbation analysis on the  Bianchi type I spacetime as done in~\cite{DeFelice:2013awa,Pookkillath:2020iqq, DeFelice:2023psw}. Another obvious direction is to study the constraints of this model by implementing the Einstein - Boltzmann code and confront with the cosmological data. It is also interesting to study the constraints of this new model with integrated Sachs-Wolfe effect and the large scale structure cross correlation data~\cite{Kimura:2011td, Kable:2021yws}, and study, if the negative cross-correlation is still existing in this model~\cite{Renk:2016olm, Giacomello:2018jfi}. Another interesting question is the consequence of the model in the non-linear region, where the velocity of the dark matter is significant, and how the screening mechanism is functioning in this model. These studies need an extensive attention, which will be carried out in other works.

\begin{acknowledgments}
The authors acknowledge the Mahidol University International Postdoctoral Fellowship Grant.
\end{acknowledgments}

\appendix
\section{Expression for $\Gamma_{cc}$, $\Gamma_{cb}$, $G_{cb}$, and $G_{cc}$} \label{apdx:Gammas_and_G}
In this appendix we give the expression for $\Gamma_{cc}$, $\Gamma_{cb}$, $G_{cb}$, and $G_{cc}$. Since these expressions are very lengthy, we write down the terms only in the leading order of the interaction term. 

\begin{equation}
    \Gamma_{cc} = 1 + \Upsilon + \mathcal{O}\big[F(Z)^2 \big] + \hdots \,
\end{equation}
where
\begin{equation}
    \Upsilon = \frac{-3 H  \dot{\phi} ^2 \ddot{\phi}  F_{,ZZ}-\dot{\phi}  \ddot{\phi} ^2 F_{,ZZ}-3 \dot{H}  \dot{\phi} ^2 F_{,Z}-9 H ^2 \dot{\phi} ^2 F_{,Z} -9 H  \dot{\phi}  \ddot{\phi}  F_{,Z}-\tfrac{d^{3}\phi}{dt^3} \dot{\phi}  F_{,Z}  -\ddot{\phi} ^2 F_{,Z}}{2 H  \rho_{\rm c} } \,.
\end{equation}

\begin{equation}
    \Gamma_{cb} \equiv \frac{ \rho_{\rm b}  \dot{\phi} ^2 F_{,Z} G_{,X} \left(3 H  \dot{\phi} +\ddot{\phi} \right)}{\rho_{\rm c}  \left(2 \alpha_{1} M_{\rm P}^2-2 M_{\rm P}^2 \dot{\phi} ^2 \ddot{\phi}  G_{,XX}-8 M_{\rm P}^2 H  \dot{\phi}  G_{,X}-4 M_{\rm P}^2 \ddot{\phi}  G_{,X}+\dot{\phi} ^4 G_{,X}^2\right)} + \mathcal{O}\big[F(Z)^2 \big] + \hdots
\end{equation}

\begin{equation}
    G_{cc}/G_{N} = G_{G_{3}} \left[ 1 + \frac{\mathcal{C}_{1}}{G_{G_{3}}}\mathcal{D} \right]  + \mathcal{O}\big[F(Z)^2 \big] + \hdots \,, \qquad G_{cb}/G_{N} = G_{G_{3}} \left[ 1 + \frac{\mathcal{C}_{2}}{G_{G_{3}}}\mathcal{D} \right]  + \mathcal{O}\big[F(Z)^2 \big] + \hdots \,,
\end{equation}
where
\begin{eqnarray}
    G_{G_{3}} \equiv \frac{2 M_{\rm P}^2 \left(\alpha_{1} - \dot{\phi} ^2 \ddot{\phi} G_{,XX} -4 H  \dot{\phi}  G_{,X}-2 \ddot{\phi}  G_{,X}\right)}{2 M_{\rm P}^2 \left(\alpha_{1}-\dot{\phi} ^2 \ddot{\phi}  G_{,XX}\right)-8 M_{\rm P}^2 H  \dot{\phi}  G_{,X}-4 M_{\rm P}^2 \ddot{\phi}  G_{,X}+\dot{\phi} ^4 G_{,X}^2} \,,
\end{eqnarray}

\begin{equation}
    \mathcal{D} \equiv \left[\alpha_{1}+\dot{\phi} ^2 \ddot{\phi}  \left(-G_{,XX}\right)-4 H  \dot{\phi}  G_{,X}-2 \ddot{\phi}  G_{,X}\right] \left[2 M_{\rm P}^2 \left(\alpha_{1}-\dot{\phi} ^2 \ddot{\phi}  G_{,XX}\right)-8 M_{\rm P}^2 H  \dot{\phi}  G_{,X}-4 M_{\rm P}^2 \ddot{\phi}  G_{,X}+\dot{\phi} ^4 G_{,X}^2\right]
\end{equation}
\begin{eqnarray}
    \mathcal{C}_{1} & \equiv & -\frac{6 F_{,Z} G_{,X}^3 \dot{\phi} ^7}{M_{\rm P}^2 \Omega_{\rm c} } - \frac{F_{,Z}
   G_{,X}^3 \dot{H}  \dot{\phi} ^7}{M_{\rm P}^2 H ^2 \Omega_{\rm c} } - \frac{G_{,X}^3 F_{,ZZ} \ddot{\phi}  \dot{\phi} ^7}{M_{\rm P}^2 H  \Omega_{\rm c} } + \frac{\alpha_{1} F_{,Z} G_{,X}^2 \dot{\phi} ^6}{M_{\rm P}^2 H  \Omega_{\rm c} } + \frac{4 F_{,Z} G_{,XX}^2 \ddot{\phi} ^2 \dot{\phi} ^6}{H  \Omega_{\rm c} } - \frac{G_{,X}^3 F_{,ZZ} \ddot{\phi} ^2 \dot{\phi} ^6}{3 M_{\rm P}^2 H ^2 \Omega_{\rm c} } \nonumber \\
   & & -\frac{3 F_{,Z} G_{,X}^3 \ddot{\phi}  \dot{\phi} ^6}{M_{\rm P}^2 H  \Omega_{\rm c} } - \frac{2 F_{,Z} G_{,X} \ddot{\phi} ^2 G_{,XXX} \dot{\phi} ^6}{H  \Omega_{\rm c} } - \frac{F_{,Z} G_{,X}^3 \tfrac{d^{3} \phi}{dt^3}  \dot{\phi} ^6}{3 M_{\rm P}^2 H ^2 \Omega_{\rm c} } + \frac{4 F_{,Z} G_{,XX}^2 \ddot{\phi} ^3 \dot{\phi} ^5}{3 H ^2 \Omega_{\rm c} } + \frac{2 G_{,X} F_{,ZZ} G_{,XX} \ddot{\phi} ^2 \dot{\phi} ^5}{H  \Omega_{\rm c} } \nonumber \\
   & & + \frac{\alpha_{1} F_{,Z} G_{,X}^2 \ddot{\phi}  \dot{\phi} ^5}{3 M_{\rm P}^2 H ^2 \Omega_{\rm c} } + \frac{20 F_{,Z} G_{,X} G_{,XX} \ddot{\phi}  \dot{\phi} ^5}{\Omega_{\rm c} } - 2 F_{,Z} G_{,X} G_{,XX} \ddot{\phi}  \dot{\phi} ^5 + \frac{2 F_{,Z} G_{,X} \dot{H}  G_{,XX} \ddot{\phi}  \dot{\phi} ^5}{H ^2 \Omega_{\rm c} } - \frac{2 F_{,Z} G_{,X} \ddot{\phi} ^3 G_{,XXX} \dot{\phi} ^5}{3 H ^2 \Omega_{\rm c} } \nonumber \\
   & & - \frac{2 F_{,Z} G_{,X} G_{,XX} \tfrac{d^{3} \phi}{dt^3}  \dot{\phi} ^5}{H  \Omega_{\rm c} } + \frac{2 G_{,X} F_{,ZZ} G_{,XX} \ddot{\phi} ^3 \dot{\phi} ^4}{3 H ^2 \Omega_{\rm c} } - 7 H  F_{,Z} G_{,X}^2 \dot{\phi} ^4 - \frac{6 H  F_{,Z} G_{,X}^2 \dot{\phi} ^4}{\Omega_{\rm c} } + \frac{50 F_{,Z} G_{,X} G_{,XX} \ddot{\phi} ^2 \dot{\phi} ^4}{3 H  \Omega_{\rm c} } \nonumber \\
   & & +\frac{8 G_{,X}^2 F_{,ZZ} \ddot{\phi}  \dot{\phi} ^4}{\Omega_{\rm c} }+G_{,X}^2 F_{,ZZ} \ddot{\phi}  \dot{\phi} ^4-\frac{6 \alpha_{1} F_{,Z} G_{,XX} \ddot{\phi}  \dot{\phi} ^4}{H  \Omega_{\rm c} }+\frac{48 N_{\rm c}^6 F_{,Z} G_{,X}^2 \rho_{{\rm c},\tilde{n}\tilde{n}\tilde{n}}\left(\tilde{n}\right) \dot{\phi} ^4}{M_{\rm P}^2 a ^{18} H  \Omega_{\rm c} ^2} + \frac{8 F_{,Z} G_{,X} G_{,XX} \ddot{\phi} ^3 \dot{\phi} ^3}{3 H ^2 \Omega_{\rm c} } \nonumber \\
   & & +\frac{20 G_{,X}^2 F_{,ZZ} \ddot{\phi} ^2 \dot{\phi} ^3}{3 H  \Omega_{\rm c} }-\frac{2 \alpha_{1} F_{,Z} G_{,XX} \ddot{\phi} ^2 \dot{\phi} ^3}{H ^2 \Omega_{\rm c} }+2 \alpha_{1} F_{,Z} G_{,X} \dot{\phi} ^3-\frac{20 \alpha_{1} F_{,Z} G_{,X} \dot{\phi} ^3}{\Omega_{\rm c} }-\frac{2 \alpha_{1} F_{,Z} G_{,X} \dot{H}  \dot{\phi} ^3}{H ^2 \Omega_{\rm c} } \nonumber \\
   & & +\frac{54 F_{,Z} G_{,X}^2 \ddot{\phi}  \dot{\phi} ^3}{\Omega_{\rm c} }-4 F_{,Z} G_{,X}^2 \ddot{\phi}  \dot{\phi} ^3+\frac{4 F_{,Z} G_{,X}^2 \dot{H}  \ddot{\phi}  \dot{\phi} ^3}{3 H ^2 \Omega_{\rm c} }-\frac{2 \alpha_{1} G_{,X} F_{,ZZ} \ddot{\phi}  \dot{\phi} ^3}{H  \Omega_{\rm c} }+\frac{16 N_{\rm c}^6 F_{,Z} G_{,X}^2 \ddot{\phi}  \rho_{{\rm c},\tilde{n}\tilde{n}\tilde{n}}\left(\tilde{n}\right) \dot{\phi} ^3}{M_{\rm P}^2 a ^{18} H ^2 \Omega_{\rm c} ^2} \nonumber \\
   & & -\frac{4 F_{,Z} G_{,X}^2 \tfrac{d^{3} \phi}{dt^3}  \dot{\phi} ^3}{3 H  \Omega_{\rm c} }+\frac{4 G_{,X}^2 F_{,ZZ} \ddot{\phi} ^3 \dot{\phi} ^2}{3 H ^2 \Omega_{\rm c} }+\frac{36 F_{,Z} G_{,X}^2 \ddot{\phi} ^2 \dot{\phi} ^2}{H  \Omega_{\rm c} }-\frac{2 \alpha_{1} G_{,X} F_{,ZZ} \ddot{\phi} ^2 \dot{\phi} ^2}{3 H ^2 \Omega_{\rm c} }+\frac{2 \alpha_{1}^2 F_{,Z} \dot{\phi} ^2}{H  \Omega_{\rm c} }-\frac{62 \alpha_{1} F_{,Z} G_{,X} \ddot{\phi}  \dot{\phi} ^2}{3 H  \Omega_{\rm c} } \nonumber \\
   & & +\frac{108 M_{\rm P}^2 H  F_{,Z} G_{,XX} \ddot{\phi}  \dot{\phi} ^2}{\Omega_{\rm c} }-\frac{96 N_{\rm c}^6 F_{,Z} G_{,XX} \ddot{\phi}  \rho_{{\rm c},\tilde{n}\tilde{n}\tilde{n}}\left(\tilde{n}\right) \dot{\phi} ^2}{a ^{18} H  \Omega_{\rm c} ^2}-\frac{2 \alpha_{1} F_{,Z} G_{,X} \tfrac{d^{3} \phi}{dt^3}  \dot{\phi} ^2}{3 H ^2 \Omega_{\rm c} }+\frac{16 F_{,Z}  G_{,X}^2 \ddot{\phi} ^3 \dot{\phi} }{3 H ^2 \Omega_{\rm c} }-\frac{4 \alpha_{1} F_{,Z} G_{,X} \ddot{\phi} ^2 \dot{\phi} }{H ^2 \Omega_{\rm c} } \nonumber \\
   & & +\frac{36 M_{\rm P}^2 F_{,Z} G_{,XX} \ddot{\phi} ^2 \dot{\phi} }{\Omega_{\rm c} }+\frac{432 M_{\rm P}^2 H ^2 F_{,Z} G_{,X} \dot{\phi} }{\Omega_{\rm c} }+\frac{2 \alpha_{1}^2 F_{,Z} \ddot{\phi}  \dot{\phi} }{3 H ^2 \Omega_{\rm c} }-\frac{32 N_{\rm c}^6 F_{,Z} G_{,XX} \ddot{\phi} ^2 \rho_{{\rm c},\tilde{n}\tilde{n}\tilde{n}}\left(\tilde{n}\right) \dot{\phi} }{a ^{18} H ^2 \Omega_{\rm c} ^2} \nonumber \\
   & & -\frac{384 N_{\rm c}^6 F_{,Z} G_{,X} \rho_{{\rm c},\tilde{n}\tilde{n}\tilde{n}}\left(\tilde{n}\right) \dot{\phi} }{a ^{18} \Omega_{\rm c} ^2}-\frac{108 \alpha_{1} M_{\rm P}^2 H  F_{,Z}}{\Omega_{\rm c} }+\frac{360 M_{\rm P}^2 H  F_{,Z} G_{,X} \ddot{\phi} }{\Omega_{\rm c} }+\frac{96 \alpha_{1} N_{\rm c}^6 F_{,Z} \rho_{{\rm c},\tilde{n}\tilde{n}\tilde{n}}\left(\tilde{n}\right)}{a ^{18} H  \Omega_{\rm c} ^2} \nonumber \\
   & & -\frac{320 N_{\rm c}^6 F_{,Z} G_{,X} \ddot{\phi}  \rho_{{\rm c},\tilde{n}\tilde{n}\tilde{n}}\left(\tilde{n}\right)}{a ^{18} H  \Omega_{\rm c} ^2}+\frac{72 M_{\rm P}^2 F_{,Z} G_{,X} \ddot{\phi} ^2}{\Omega_{\rm c}  \dot{\phi} }-\frac{36 \alpha_{1} M_{\rm P}^2 F_{,Z} \ddot{\phi} }{\Omega_{\rm c}  \dot{\phi} }-\frac{64 N_{\rm c}^6 F_{,Z} G_{,X} \ddot{\phi} ^2 \rho_{{\rm c},\tilde{n}\tilde{n}\tilde{n}}\left(\tilde{n}\right)}{a ^{18} H ^2 \Omega_{\rm c} ^2 \dot{\phi} }+\frac{32 \alpha_{1} N_{\rm c}^6 F_{,Z} \ddot{\phi}  \rho_{{\rm c},\tilde{n}\tilde{n}\tilde{n}}\left(\tilde{n}\right)}{a ^{18} H ^2 \Omega_{\rm c} ^2 \dot{\phi} } \nonumber \\
\end{eqnarray}
\begin{eqnarray}
    \mathcal{C}_{2} & \equiv & -\frac{6 F_{,Z} G_{,X}^3 \dot{\phi} ^7}{M_{\rm P}^2 \Omega_{\rm c} }-\frac{F_{,Z} G_{,X}^3 \dot{H}  \dot{\phi} ^7}{M_{\rm P}^2 H ^2 \Omega_{\rm c} }-\frac{G_{,X}^3 F_{,ZZ} \ddot{\phi}  \dot{\phi} ^7}{M_{\rm P}^2 H  \Omega_{\rm c} }+\frac{\alpha_{1} F_{,Z} G_{,X}^2 \dot{\phi} ^6}{M_{\rm P}^2 H  \Omega_{\rm c} }+\frac{4 F_{,Z} G_{,XX}^2 \ddot{\phi} ^2 \dot{\phi} ^6}{H \Omega_{\rm c} }-\frac{G_{,X}^3 F_{,ZZ} \ddot{\phi} ^2 \dot{\phi} ^6}{3 M_{\rm P}^2 H ^2 \Omega_{\rm c} } \nonumber \\
    & & -\frac{3 F_{,Z} G_{,X}^3 \ddot{\phi}  \dot{\phi} ^6}{M_{\rm P}^2 H  \Omega_{\rm c} }-\frac{2 F_{,Z} G_{,X} \ddot{\phi} ^2 G_{,XXX} \dot{\phi} ^6}{H  \Omega_{\rm c} }-\frac{F_{,Z} G_{,X}^3 \tfrac{d^{3} \phi}{dt^{3}}  \dot{\phi} ^6}{3 M_{\rm P}^2 H ^2 \Omega_{\rm c} }+\frac{4 F_{,Z} G_{,XX}^2 \ddot{\phi} ^3 \dot{\phi} ^5}{3 H ^2 \Omega_{\rm c} }+\frac{2 G_{,X} F_{,ZZ} G_{,XX} \ddot{\phi} ^2 \dot{\phi} ^5}{H  \Omega_{\rm c} } \nonumber \\
    & & +\frac{\alpha_{1} F_{,Z} G_{,X}^2 \ddot{\phi}  \dot{\phi} ^5}{3 M_{\rm P}^2 H ^2 \Omega_{\rm c} }+\frac{20 F_{,Z} G_{,X} G_{,XX} \ddot{\phi}  \dot{\phi} ^5}{\Omega_{\rm c} }-2 F_{,Z} G_{,X} G_{,XX} \ddot{\phi}  \dot{\phi} ^5+\frac{2 F_{,Z} G_{,X} \dot{H}  G_{,XX} \ddot{\phi}  \dot{\phi} ^5}{H ^2 \Omega_{\rm c} }-\frac{2 F_{,Z} G_{,X} \ddot{\phi} ^3 G_{,XXX} \dot{\phi} ^5}{3 H ^2 \Omega_{\rm c} } \nonumber \\
    & & -\frac{2 F_{,Z} G_{,X} G_{,XX} \tfrac{d^{3} \phi}{dt^{3}}  \dot{\phi} ^5}{H  \Omega_{\rm c} }+\frac{2 G_{,X} F_{,ZZ} G_{,XX} \ddot{\phi} ^3 \dot{\phi} ^4}{3 H ^2 \Omega_{\rm c} }-7 H  F_{,Z} G_{,Z}^2 \dot{\phi} ^4+\frac{48 H  F_{,Z} G_{,X}^2 \dot{\phi} ^4}{\Omega_{\rm c} }+\frac{50 F_{,Z} G_{,X} G_{,XX} \ddot{\phi} ^2 \dot{\phi} ^4}{3 H  \Omega_{\rm c} } \nonumber \\
    & & +\frac{8 G_{,X}^2 F_{,ZZ} \ddot{\phi}  \dot{\phi} ^4}{\Omega_{\rm c} }+G_{,Z}^2 F_{,ZZ} \ddot{\phi}  \dot{\phi} ^4-\frac{6 \alpha_{1} F_{,Z} G_{,XX} \ddot{\phi}  \dot{\phi} ^4}{H  \Omega_{\rm c} }+\frac{6 N_{\rm b}^3 F_{,Z} G_{,X}^2 \rho_{,n_{b}n_{b}n_{b}}\left(n_{\rm b}\right) \dot{\phi} ^4}{M_{\rm P}^2 a ^9 H  \Omega_{\rm b}  \Omega_{\rm c} }+\frac{8 F_{,Z} G_{,X} G_{,XX} \ddot{\phi} ^3 \dot{\phi} ^3}{3 H ^2 \Omega_{\rm c} } \nonumber \\
    & & +\frac{20 G_{,X}^2 F_{,ZZ} \ddot{\phi} ^2 \dot{\phi} ^3}{3 H  \Omega_{\rm c} }-\frac{2 \alpha_{1} F_{,Z} G_{,XX} \ddot{\phi} ^2 \dot{\phi} ^3}{H ^2 \Omega_{\rm c} }+2 \alpha_{1} F_{,Z} G_{,X} \dot{\phi} ^3-\frac{20 \alpha_{1} F_{,Z} G_{,X} \dot{\phi} ^3}{\Omega_{\rm c} }-\frac{2 \alpha_{1} F_{,Z} G_{,X} \dot{H}  \dot{\phi} ^3}{H ^2 \Omega_{\rm c} }+\frac{72 F_{,Z} G_{,X}^2 \ddot{\phi}  \dot{\phi} ^3}{\Omega_{\rm c} } \nonumber \\
    & & -4 F_{,Z} G_{,X}^2 \ddot{\phi}  \dot{\phi} ^3+\frac{4 F_{,Z} G_{,X}^2 \dot{H}  \ddot{\phi}  \dot{\phi} ^3}{3 H ^2 \Omega_{\rm c} }-\frac{2 \alpha_{1} G_{,X} F_{,ZZ} \ddot{\phi}  \dot{\phi} ^3}{H  \Omega_{\rm c} }+\frac{2 N_{\rm b}^3 F_{,Z} G_{,X}^2 \ddot{\phi}  \rho_{,n_{b}n_{b}n_{b}}\left(n_{\rm b}\right) \dot{\phi} ^3}{M_{\rm P}^2 a ^9 H ^2 \Omega_{\rm b}  \Omega_{\rm c} }-\frac{4 F_{,Z} G_{,X}^2 \tfrac{d^{3} \phi}{dt^{3}}  \dot{\phi} ^3}{3 H  \Omega_{\rm c} } \nonumber \\
    & & +\frac{4 G_{,X}^2 F_{,ZZ} \ddot{\phi} ^3 \dot{\phi} ^2}{3 H ^2 \Omega_{\rm c} }+\frac{36 F_{,Z} G_{,X}^2 \ddot{\phi} ^2 \dot{\phi} ^2}{H  \Omega_{\rm c} }-\frac{2 \alpha_{1} G_{,X} F_{,ZZ} \ddot{\phi} ^2 \dot{\phi} ^2}{3 H ^2 \Omega_{\rm c} }+\frac{2 \alpha_{1}^2 F_{,Z} \dot{\phi} ^2}{H  \Omega_{\rm c} }-\frac{62 \alpha_{1} F_{,Z} G_{,X} \ddot{\phi}  \dot{\phi} ^2}{3 H  \Omega_{\rm c} } \nonumber \\
    & & -\frac{12 N_{\rm b}^3 F_{,Z} G_{,XX} \ddot{\phi}  \rho_{,n_{b}n_{b}n_{b}}\left(n_{\rm b}\right) \dot{\phi} ^2}{a ^9 H  \Omega_{\rm b}  \Omega_{\rm c} }-\frac{2 \alpha_{1} F_{,Z} G_{,X} \tfrac{d^{3} \phi}{dt^{3}}  \dot{\phi} ^2}{3 H ^2 \Omega_{\rm c} }+\frac{16 F_{,Z} G_{,X}^2 \ddot{\phi} ^3 \dot{\phi} }{3 H ^2 \Omega_{\rm c} }-\frac{4 \alpha_{1} F_{,Z} G_{,X} \ddot{\phi} ^2 \dot{\phi} }{H ^2 \Omega_{\rm c} }+\frac{2 \alpha_{1}^2 F_{,Z} \ddot{\phi}  \dot{\phi} }{3 H ^2 \Omega_{\rm c} } \nonumber \\
    & & -\frac{4 N_{\rm b}^3 F_{,Z} G_{,XX} \ddot{\phi} ^2 \rho_{,n_{b}n_{b}n_{b}}\left(n_{\rm b}\right) \dot{\phi} }{a ^9 H ^2 \Omega_{\rm b}  \Omega_{\rm c} }-\frac{48 N_{\rm b}^3 F_{,Z} G_{,X} \rho_{,n_{b}n_{b}n_{b}}\left(n_{\rm b}\right) \dot{\phi} }{a ^9 \Omega_{\rm b}  \Omega_{\rm c} }+\frac{12 \alpha_{1} N_{\rm b}^3 F_{,Z} \rho_{,n_{b}n_{b}n_{b}}\left(n_{\rm b}\right)}{a ^9 H  \Omega_{\rm b}  \Omega_{\rm c} }\nonumber \\
    & & -\frac{40 N_{\rm b}^3 F_{,Z} G_{,X} \ddot{\phi}  \rho_{,n_{b}n_{b}n_{b}}\left(n_{\rm b}\right)}{a ^9 H  \Omega_{\rm b}  \Omega_{\rm c} }-\frac{8 N_{\rm b}^3 F_{,Z} G_{,X} \ddot{\phi} ^2 \rho_{,n_{b}n_{b}n_{b}}\left(n_{\rm b}\right)}{a ^9 H ^2 \Omega_{\rm b}  \Omega_{\rm c}  \dot{\phi} }+\frac{4 \alpha_{1} N_{\rm b}^3 F_{,Z} \ddot{\phi}  \rho_{,n_{b}n_{b}n_{b}}\left(n_{\rm b}\right)}{a ^9 H ^2 \Omega_{\rm b}  \Omega_{\rm c}  \dot{\phi} }
\end{eqnarray}

\section{Behaviour of $G_{cc}$ and $G_{bb}$ along the redshift} \label{apdx:geff}
\begin{figure}[h]
    \centering
    \includegraphics[]{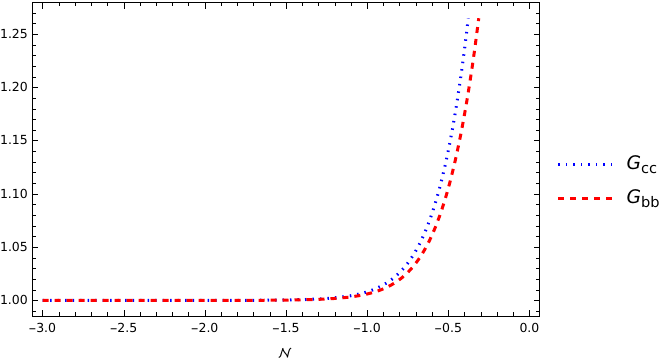}
    \caption{The behaviour of effective gravitational coupling to the cold dark matter and the baryon $G_{cc}$ and $G_{bb}$ respectively. }
    \label{fig:Gcc_Gbb}
\end{figure}

In this appendix we show the behaviour of the $G_{cc}/G_{N}$ and $G_{bb}/G_{N}$ along the variable $\mathcal{N} = \ln(a)$, Fig.~\ref{fig:Gcc_Gbb}. As we can see that there is enhancement in the ratio between $G_{cc} (G_{bb})$ and $G_{N}$ in the dark energy domination era.

\bibliographystyle{JHEP}
\bibliography{refs}

\end{document}